\documentclass[12pt]{article}
\usepackage{graphicx}

\def\CP{$ C \! P$ }

\def\Cseven{$C^{eff}_7$~}
\def\Csevenp{$C^{eff}_7$}
\def\Ceight{$C^{eff}_8$~}
\def\Ceightp{$C^{eff}_8$}
\def\Cnine{$C^{eff}_9$~}

\def\Cten{$C^{eff}_{10}$~}

\def\ra{\rightarrow}

\def\sbabar{\mbox{{\normalsize \sl B}\hspace{-0.4em} {\small \sl A}\hspace{-0.03em}{\normalsize \sl B}\hspace{-0.4em} {\small \sl A\hspace{-0.02em}R}}}

\newcommand\T{\rule{0pt}{2.6ex}}       
\newcommand\TT{\rule{0pt}{3.0ex}}       
\newcommand\B{\rule[-1.2ex]{0pt}{0pt}} 

\newcommand\pubnumber{}
\newcommand\pubdate{\today}

\def\Title#1{\begin{center} {\Large #1 } \end{center}}
\def\Author#1{\begin{center}{ \sc #1} \end{center}}
\def\Address#1{\begin{center}{ \it #1} \end{center}}

\newcommand\pubblock{\rightline{\begin{tabular}{l} \pubnumber\\
         \pubdate  \end{tabular}}}
\newenvironment{Abstract}{\begin{quotation}  }{\end{quotation}}
\newenvironment{Presented}{\begin{quotation} \begin{center} 
             PRESENTED AT\end{center}\bigskip 
      \begin{center}\begin{large}}{\end{large}\end{center} \end{quotation}}
\def\Acknowledgements{\bigskip  \bigskip \begin{center} \begin{large}
             \bf ACKNOWLEDGEMENTS \end{large}\end{center}}

\def\beq{\begin{equation}}
\def\eeq#1{\label{#1}\end{equation}}
\def\eeqn{\end{equation}}

\def\beqa{\begin{eqnarray}}
\def\eeqa#1{\label{#1}\end{eqnarray}}
\def\eeqan{\end{eqnarray}}

\let\bar=\overbar

\def\Dslash{\not{\hbox{\kern-4pt $D$}}}
\def\dslash{\not{\hbox{\kern-2pt $\del$}}}

\def\BR{\mbox{\rm BR}}

\def\msb{{\bar{\ssstyle M \kern -1pt S}}}

\begin{document}
\begin{titlepage}
\pubblock

\vfill
\Title{Radiative Penguin Decays at $e^+ e^-$ Colliders}
\vfill
\Author{ Gerald Eigen  \footnote{Work supported by the Norwegian Research Council.} \\
(representing the \sbabar\ collaboration)}
\Address{Department of Physics\\
University of Bergen, N-5007 Bergen, NORWAY 
}

\vfill
\begin{Abstract}
In this review, the most recent results of the radiative decays $B \ra X_s \gamma$, $B \ra K^{(*)} \ell^+ \ell^-$ and $B \ra \pi/\eta  \ell^+ \ell^-$ at $e^+e^-$ colliders are discussed. The new, most precise \CP asymmetry measurements in $B \ra X_s \gamma$ from \sbabar\ are presented together with branching fractions and photon energy moments. 
For $B \ra K^{(*)} \ell^+ \ell^-$ modes, $B$ factory results on partial branching fractions, rate asymmetries and angular observables are combined with measurements from CDF and the LHC experiments. The first branching fraction upper limits for $B \ra \eta  \ell^+ \ell^-$ are shown along with updated upper limits of $B \ra \pi  \ell^+ \ell^-$ branching fractions.

\end{Abstract}
\vfill
\begin{Presented}
Flavor Physics and \CP Violation\\
Buzios, Rio, Brasil,  May 19--24, 2013
\end{Presented}
\vfill
\end{titlepage}
\def\thefootnote{\fnsymbol{footnote}}
\setcounter{footnote}{0}

\section{Introduction}

The decays $B \ra X_{s(d)} \gamma$ and $B \ra X_{s(d)} \ell^+ \ell^-$, where $\ell^+ \ell^-$ is $e^+ e^-$ or $\mu^+ \mu^-$, are flavor-changing neutral-current processes that are forbidden in the Standard Model (SM) at tree level.  They occur in higher-order processes and are described by an effective Hamiltonian that factorizes short-distance contributions in terms of scale-dependent Wilson coefficients $C_i(\mu)$ from long-distance effects expressed by local four-fermion operators ${\cal O}_i$ that define hadronic matrix elements, 
\begin{equation}
H_{eff}=\frac{4 G_F}{\sqrt{2}} \sum_i C_i(\mu) {\cal O}_i.
\end{equation} 
\noindent
While Wilson coefficients are calculable perturbatively, the calculation of the hadronic matrix elements requires non-perturbative methods such as the heavy quark expansion~\cite{wilson, isgur, georgi, grinstein}.

\begin{figure}[h]
\centering
\vskip -0.4cm
\includegraphics[width=63mm]{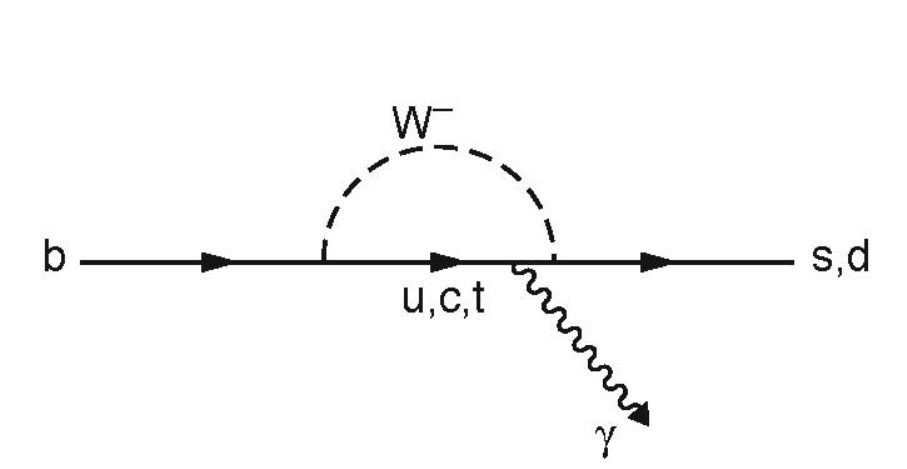}
\includegraphics[width=88mm]{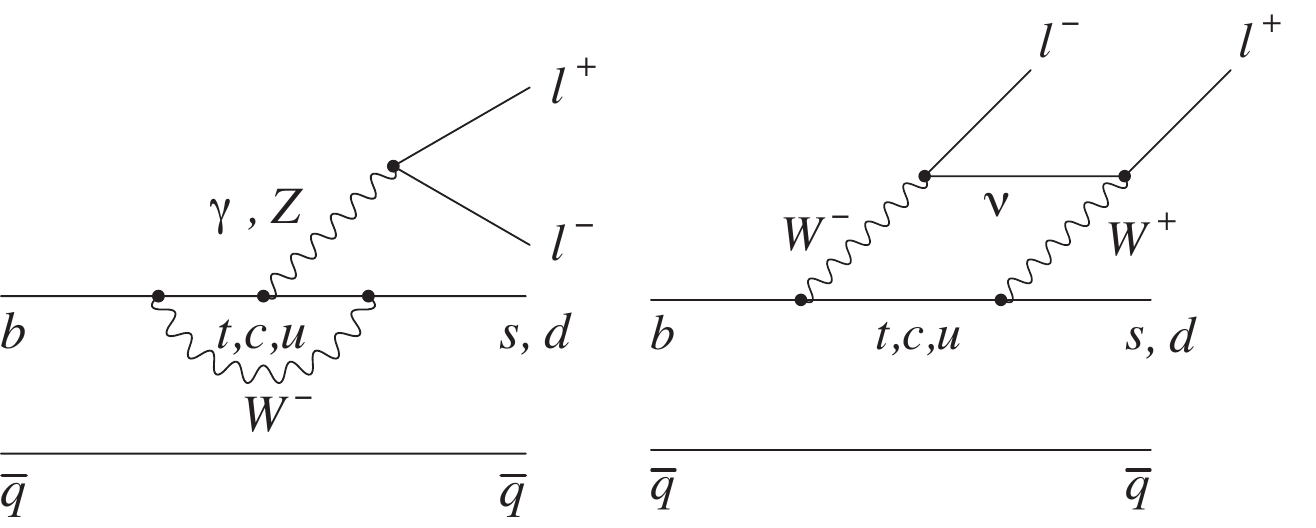}
\caption{Lowest-order diagrams for  $B \ra X_{s(d)} \gamma$ (left) and $B \ra X_{s(d)} \ell^+ \ell^-$ (middle, right).}
 \label{fig:penguin}
\end{figure}

\begin{figure}[h]
\centering
\includegraphics[width=120mm]{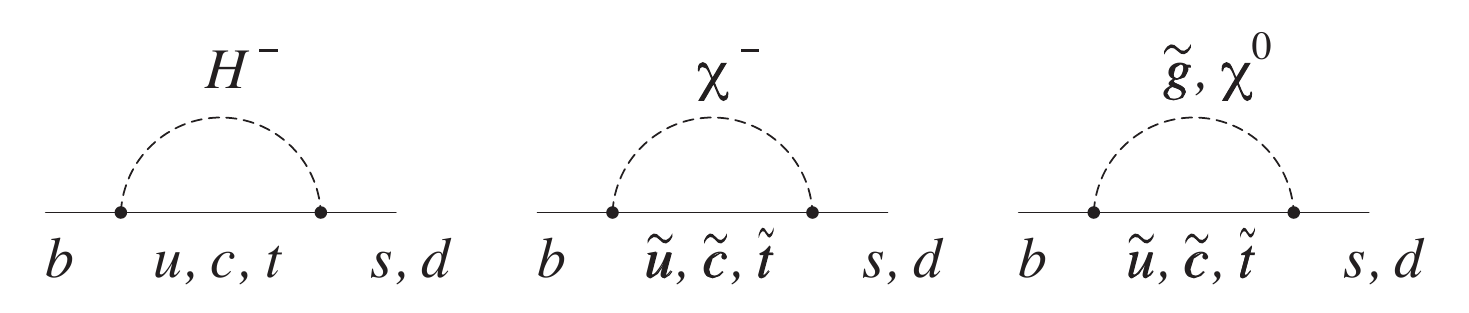}
\caption{Examples of new physiccs processes.}
 \label{fig:np}
\end{figure}

Figure~\ref{fig:penguin} shows the lowest order diagrams. In the $B \ra X_{s(d)} \gamma$ decay, the electromagnetic penguin loop dominates. The short-distance part is expressed by the effective Wilson coefficient \Csevenp.  Through operator mixing at higher orders, the chromo-magnetic penguin enters whose short distance part is parameterized by \Ceightp. In $B \ra X_{s(d)} \ell^+ \ell^-$ modes, the $Z$ penguin and the $WW$ box diagram contribute in addition. Their short-distance parts are parametrized in terms of  \Cnine (vector current part) and \Cten (axial-vector current part). Physics beyond the SM introduces new loops and box diagrams with new particles ({\it e.g.} charged Higgs boson, supersymmetric particles) as shown in Fig.~\ref{fig:np}. Such contributions modify the Wilson coefficients and may introduce new diagrams with scalar and pseudoscalar current interactions and in turn new Wilson coefficients, $C_S$ and $C_P$. To determine \Csevenp, \Ceight, \Cnine and \Cten precisely, we need to measure many observables in several radiative decays. These rare decays can potentially probe new physics at a scale of a few TeV. 

We present herein new \sbabar\ measurements of the direct \CP  asymmetry in $B \ra X_s \gamma$ using a semi-inclusive analysis and determine the ratio of Wilson coefficients ${\cal I}m(C^{eff}_8/C^{eff}_7)$. In addition, we summarize the status of partial branching fractions, photon energy spectra, photon energy moments and \CP asymmetries from $e^+ e^-$ colliders for fully inclusive and semi-inclusive $B \ra X_s \gamma$ analyses.  We determine  the $b$ quark mass  $m_b$ and its kinetic energy $\mu^2_\pi$ in the kinetic and shape function models. We review the status of branching fractions, rate asymmetries and angular observables for $B \ra K^{(*)} \ell^+ \ell^-$ modes. Finally, we present a new \sbabar\ search for $B \ra \pi \ell^+ \ell^-$ modes and a first search for $B \ra \eta \ell^+ \ell^-$modes. \sbabar\ performs all analyses  blinded.

\section{Study of $B \ra X_s \gamma$}

In the SM, the $B \ra X_s \gamma$ branching fraction is calculated at next-to-next-to-leading order (up to four loops) yielding ${\cal B}(B \ra X_S \gamma)=(3.14\pm0.22)\times 10^{-4} $ for photon energies $E^*_\gamma >1.6\rm ~GeV$~\cite{misiak07, misiak07a, misiak13}. For larger minimum values of $E^*_\gamma$, the prediction depends on the shape of the $E^*_\gamma$ spectrum, which is modeled in terms of a shape function that depends on the Fermi motion of the $b$ quark inside the $B$ meson and thus on the $b$ quark mass. Since the shape function is expected to be similar to that used to determine the lepton-energy spectrum in $B \ra X_u \ell \nu$, precision measurements of the $E^*_\gamma$ spectrum are helpful for the determination of $V_{ub}$. The measurement of ${\cal B}(B \ra X_s \gamma)$ provides constraints on the charged Higgs mass $m_{H^\pm}$.

Experimentally, the challenge is to extract the $E_\gamma^*$ signal from photon background copiously produced in $\pi^0$ decays in $q \bar q$  continuum\footnote{q refers to $u, d, s$ and $c$} and $B \bar B$ processes that increases exponentially with smaller photon energy. We use three different strategies to suppress these backgrounds: i) an inclusive analysis with a lepton tag, ii) a semi-inclusive analysis and iii) an inclusive analysis with a fully reconstructed $B$ meson. Herein, we present results of the first two strategies.

\subsection{Fully Inclusive $B \ra X_s \gamma$ Analysis}

\begin{figure}[h]
\centering
\includegraphics[width=66mm]{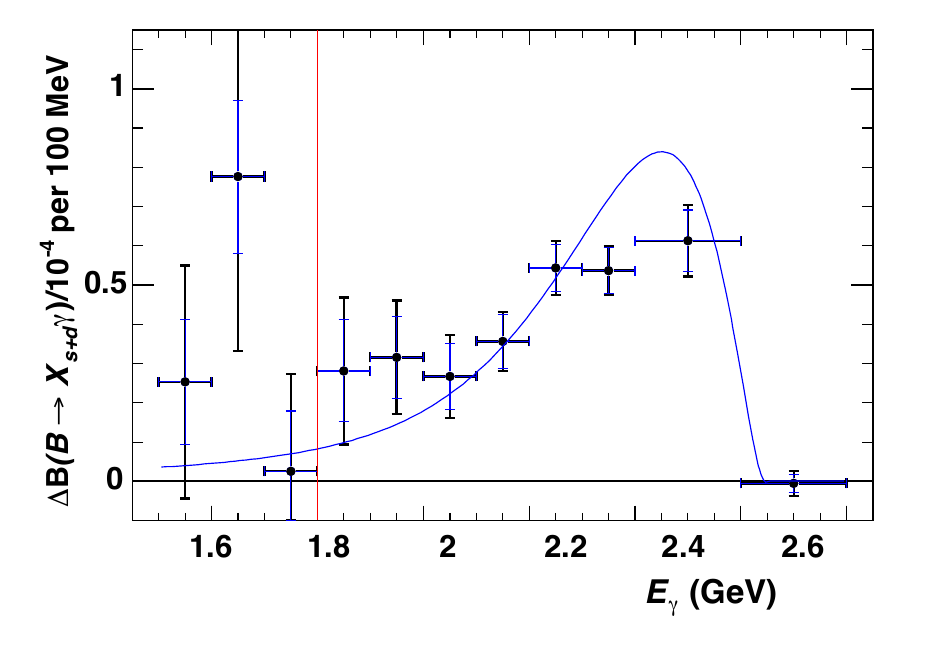}
\includegraphics[width=85mm]{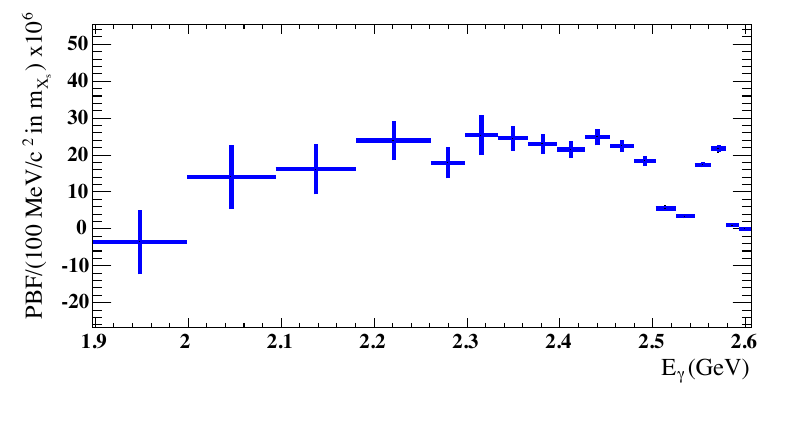}
\caption{Partial branching fraction versus $E^*_\gamma$ measured in a fully inclusive analysis (left) and for the sum of exclusive modes (right). 
Error bars (left) show statistical and total uncertainties.
The solid curve shows a prediction for the kinetic scheme with HFAG averages~\cite{hfag}. The vertical bar separates signal from the control region. Errors bars (right) show total uncertainties.}
 \label{fig:bsg}
\end{figure}

Using a sample of $384 \times 10^6~ B \bar B$ events, \sbabar\ measured total and partial branching fractions, photon energy moments and the $B \ra X_{s+d} \gamma$ \CP asymmetry in a fully inclusive analysis~\cite{babar12, babar12a}. To suppress $e^+ e^- \ra q \bar q$ continuum and $B \bar B$ backgrounds, we tag the recoiling $B$ meson in semileptonic decays and use optimized $\pi^0$ and $\eta$ vetoes, missing energy requirements and the output of two neural networks (NN). For a signal efficiency of $2.5\%$, the efficiency for accepting continuum ($B \bar B$) background is reduced to $5 \times 10^{-6}~ (1.3 \times 10^{-4})$. We estimate the residual continuum background by studying data taken 40~MeV below the $\Upsilon(4S)$ peak. Figure~\ref{fig:bsg} (left) shows the  $B \ra X_s \gamma$ partial branching fraction after background subtraction and corrections for efficiency, resolution effects and Doppler smearing. For comparison, we show the predicted $E^*_\gamma$ spectrum in the kinetic scheme~\cite{benson, neubert} using HFAG world averages~\cite{hfag} for the shape function parameters. For $E^*_\gamma > \rm 1.8~GeV$, \sbabar\ measures a total branching fraction of ${\cal B}(B \ra X_S \gamma)=(3.21\pm 0.15_{stat} \pm 0.29_{sys} \pm 0.08_{model}) \times 10^{-4} $, where uncertainties are statistical, systematic and from model dependence, respectively. This is
in good agreement with previous measurements~\cite{babar5, belle9, cleo1}.  After extrapolation
to  $E_\gamma >1.6 \rm ~GeV$, the branching fraction increases to ${\cal B}(B \ra X_S \gamma)=(3.31\pm 0.16_{stat} \pm 0.30_{sys} \pm 0.09_{model}) \times 10^{-4} $, which is still in good agreement with the SM prediction. We use this result to constrain new physics in the type II two-Higgs doublet model~\cite{misiak07,haisch, ciuchini} excluding $m_{H^\pm} \rm < 327~GeV/c^2$  at $95\%$ confidence level ($CL$) independent of $\tan \beta$. Recent \sbabar\ results on ${\cal B}(B \ra D^{(*)} \tau \nu)$, however, are in conflict with both the SM and the type II Higgs doublet model at the $3\sigma$ level~\cite{babar13, babar13a}.

For $E^*_\gamma >1.8~\rm GeV$, \sbabar\ measured energy moments of
$ \langle E_\gamma \rangle = (2.267 \pm 0.019_{stat} \pm 0.032_{sys} \pm 0.003_{mod}) ~\rm GeV $ and 
$\langle (E_\gamma  - \langle E_\gamma \rangle )^2 \rangle =(0.0484 \pm 0.0053_{stat} \pm 0.0077_{sys} \pm 0.0005_{mod})~\rm GeV^2$ that are consistent with previous results~\cite{babar5, belle9, cleo1}. where uncertainties are statistical, systematic and from model dependence, respectively.

\subsection{Semi-Inclusive $B \ra X_s \gamma$ Analysis}

Using $471~ B\bar B$ events in a semi-inclusive analysis, we combine 38 exclusive $B \ra X_s \gamma$ final states containing a $K^+$\footnote{Charge conjugation is implied throughout the article unless stated otherwise} or $K^0_S$ and up to four pions with at most two $\pi^0$s, $K^+K^-$ with up to one pion,  or up to one $\eta$ with up to two pions
~\cite{babar12b}. We reconstruct the hadronic mass $m_{X_s}$ in $\rm 100~MeV/c^2$ bins and calculate the photon energy by $E_\gamma =\frac{m^2_B - m^2_{X_s}}{2 m_B}$. Figure~\ref{fig:bsg} (right) shows the partial branching fraction versus  $E^*_\gamma$.  Summing the partial branching fraction over all $m_{X_S}$ bins yields ${\cal B}(B \ra X_s \gamma)=(3.29\pm 0.19_{stat}\pm 0.48_{sys}) \times 10^{-4}$ for $E_\gamma > \rm 1.9~GeV$, which is in good agreement with the results of the inclusive analysis. 
We also measure the mean and variance of the photon energy spectrum, $ \langle E_\gamma \rangle = (2.346 \pm 0.018^{+0.027}_{-0.022} )~\rm GeV $ and 
$\langle (E_\gamma  - \langle E_\gamma \rangle )^2 \rangle =(0.0211\pm 0.0057^{+0.0055}_{-0.0069})~\rm GeV^2$ for $E_\gamma >1.9~\rm GeV$. These results agree with the measurements of the inclusive analysis after increasing the minimum $E^*_\gamma$ selection to 1.9~GeV. Note that  $\langle E_\gamma \rangle$ ($\langle (E_\gamma  - \langle E_\gamma \rangle )^2 $) increases (decreases) with a larger minimum $E^*_\gamma$ selection. 
From a fit to the photon energy spectrum, we can extract the $b$ quark mass and its kinetic energy.
Table ~\ref{tab:bsg} summarizes the results of $m_b$ and $\mu^2_\pi$ for fits to the $E^*_\gamma$ spectrum in the kinetic scheme~\cite{benson} and shape function scheme~\cite{lange}.

\vskip 0.2cm

\begin{table}
\centering
\caption{Determination of $m_b$ and $\mu^2_\pi$ in the kinetic-scheme~\cite{benson} and shape function scheme~\cite{lange} using the semi-inclusive analysis in comparison to the world average~\cite{hfag}.}
\medskip
{\footnotesize
\begin{tabular}{|l|c|c|c|c|}
\hline \hline \TT
& \sbabar & \sbabar & world average & world average \\
& kinetic scheme & shape function scheme & kinetic scheme & shape function scheme  \T \B
\\ \hline
$m_b~\rm  [GeV/c^2]$ & $4.568^{+0.038}_{-0.036}$ & $4.579^{+0.032}_{-0.029}$ & $4.560\pm0.023$ & $4.588\pm 0.025$ \T \B   \\
$\mu^2_\pi~ \rm [GeV^2]$ & $0.450\pm 0.054$ & $0.257^{+0.034}_{-0.039} $& $0.453\pm 0.036$ & $0.189^{+0.046}_{-0.057} $ \T \B
   \\
\hline \hline                        
\end{tabular}
}
\label{tab:bsg}
\end{table}

\begin{figure}[h]
\centering
\includegraphics[width=120mm]{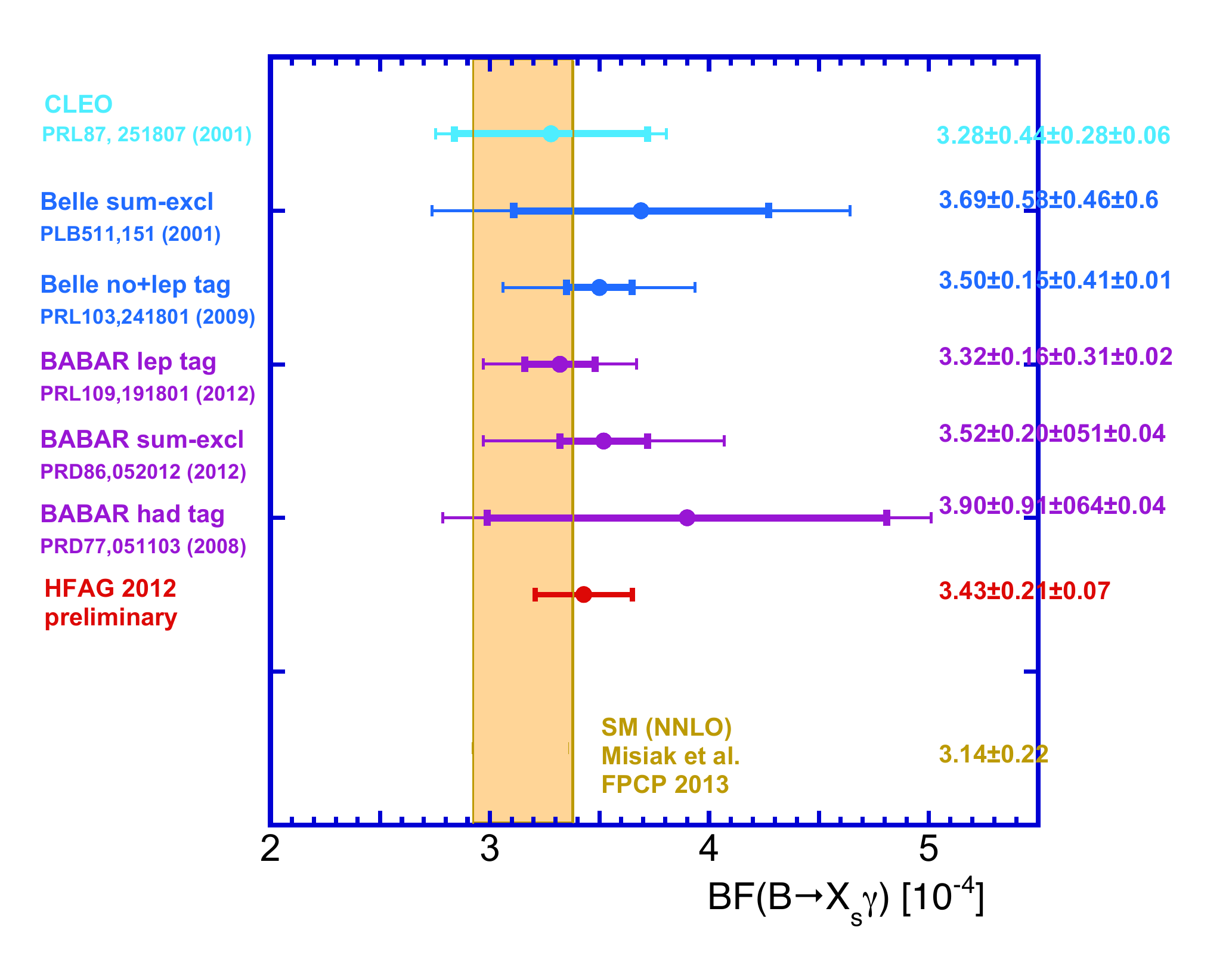}
\caption{Comparison of ${\cal B}(B\ra X_s \gamma)$ measurements from \sbabar~\cite{babar12, babar12b, babar8}, Belle~\cite{belle9, belle1} and CLEO~\cite{cleo1} to the SM prediction~\cite{misiak13} after extrapolation to $E^*_\gamma > 1.6 ~\rm GeV$. }
 \label{fig:bfbsg}
\end{figure}

Figure~\ref{fig:bfbsg} shows a comparison of all  $B\ra X_s \gamma$ total branching fraction measurements after extrapolating them to a $E^*_\gamma >1.6~\rm GeV$ selection. In addition, the HFAG average~\cite{hfag} and the SM prediction~\cite{misiak13} are depicted. All ${\cal B}(B\ra X_s \gamma)$ measurements are in good agreement with each other and with the SM prediction.

\subsection{Direct \CP Asymmetry} 

For the sum of exclusive modes, the direct \CP asymmetry is defined by

\begin{equation}
{\cal A}_{C\!P} (B\ra X_s \gamma) \equiv \frac{{\cal B} (\bar B \ra \bar X_s \gamma) - {\cal B} (B \ra X_s \gamma) } {{\cal B}(\bar B \ra \bar X_s \gamma) + {\cal B} (B \ra X_s \gamma)}.
\end{equation}

The present world average of ${\cal A}_{C\! P} = (-0.8\pm 2.9)\%$ is in good agreement with the SM prediction of $-0.6\% < {\cal A}_{C \!P} < 2.8\%$ at $95\%$ CL~\cite{benzke}. The presently large uncertainties still allow for new physics contributions, which modify  $C_7^{eff}$. Particularly, the \CP asymmetry difference between $B^+$ and $B^0$ decays, $\Delta {\cal A}_{C \! P}(B \ra X_s \gamma) = {\cal A}_{C \! P}(B^+ \ra X^+_s \gamma) -{\cal A}_{C \! P} (B^0 \ra X^0_s \gamma) $, is very sensitive to new physics since it is caused by interference between the electromagnetic and the chromo-magnetic penguin diagrams in which the latter enters through higher-order corrections. Calculations yield~\cite{benzke}

\begin{equation}
\Delta {\cal A}_{C \! P}(B \ra X_s \gamma) \simeq 4 \pi^2 \alpha_s \frac{\bar \Lambda_{78}}{m_b} {\cal I}m \frac{C_8^{eff}}{C_7^{eff}} \simeq 0.12 \frac{\bar \Lambda_{78}}{100~\rm MeV}{\cal I}m \frac{C_8^{eff}}{C_7^{eff}},
\end{equation}
\noindent
where  $\bar \Lambda_{78} $ is the hadronic matrix element of the ${\cal O}_7-{\cal O}_8$ interference, predicted to lie in the range 17~MeV $ <\bar \Lambda_{78} <$ 190~MeV.
In the SM, $\Delta {\cal A}_{C \! P} (B \ra X_s \gamma)$ vanishes since  $C_7^{eff}$ and $C_8^{eff}$ are real. 

In a sample of $471 ~B \bar B$ events, \sbabar\ studied ${\cal A_{C \! P}}$ and $\Delta {\cal A}_{C \! P}$ in a semi-inclusive analysis using ten $B^+$ and six $B^0$ exclusive final states.\footnote{$B^+ \ra K^0_S \pi^+ \gamma, K^+ \pi^0 \gamma, K^+ \pi^+ \pi^- \gamma, K^0_S \pi^+ \pi^0 \gamma,
K^+ \pi^0 \pi^0 \gamma, K^0_S \pi^+ \pi^- \pi^+, K^+ \pi^+ \pi^- \pi^0, K^0_S \pi^+ \pi^0 \pi^0,$ $ K^+ \eta \gamma, K^+ K^+ K^- \gamma$ and $B^0 \ra K^+ \pi^- \gamma, K^+ \pi^- \pi^0 \gamma, K^+ \pi^+ \pi^- \pi^- \gamma, K^+ \pi^- \pi^0 \pi^0 \gamma, K^+ \pi^- \eta \gamma,
3K^\pm  \pi^- \gamma$.} We maximize the signal extraction using a bagged decision tree with six input variables. This improves the efficiency considerably with respect to the standard $\Delta E=E^*_B-E^*_{beam}$ selection, where $E^*_{beam}$ and $E_B^*$ are the beam energy and $B$ meson energy in the center-of-mass frame, respectively. To remove continuum background, we train a separate bagged decision tree using event shape variables. We perform an $X_s$ mass-dependent optimization with loosely identified pions and kaons using the sensitivity $ S/\sqrt{S+B}$ where $S ~(B)$ is the signal (background) yield. To extract ${\cal A}_{C \! P}$, we fit the beam energy-constrained mass $m_{ES}=\sqrt{E^{*2}_{beam}-p^{*2}_B}$\footnote{$p^*_B$ is the $B$ momentum in the center-of-mass frame. }  simultaneously for $\bar B$-tagged and $B$-tagged samples.
After correcting the raw  ${\cal A}_{C \! P}$ for detector bias determined from the $m_{ES}$ sideband below the signal region, we measure ${\cal A}_{C \! P}(B \ra X_s \gamma)= (1.73\pm1.93_{stat} \pm 1.02_{sys})\%$, which agrees well with the SM prediction. This new measurement has the smallest uncertainty. From a simultaneous fit to $B^+$ and $B^0$ samples, we measure $\Delta {\cal A}_{C \! P}(B \ra X_s \gamma)= (4.97\pm3.90_{stat} \pm 1.45_{sys})\%$ from which we obtain the constraint  $-1.64 <  {\cal I}m(C^{eff}_8/C^{eff}_7) < 6.52 $ at $90\%$ CL. Note, this is the first $\Delta {\cal A}_{C \! P}$ measurement and first constraint on ${\cal I}m(C^{eff}_8/C^{eff}_7)$.

\begin{figure}[h]
\centering
\includegraphics[width=80mm]{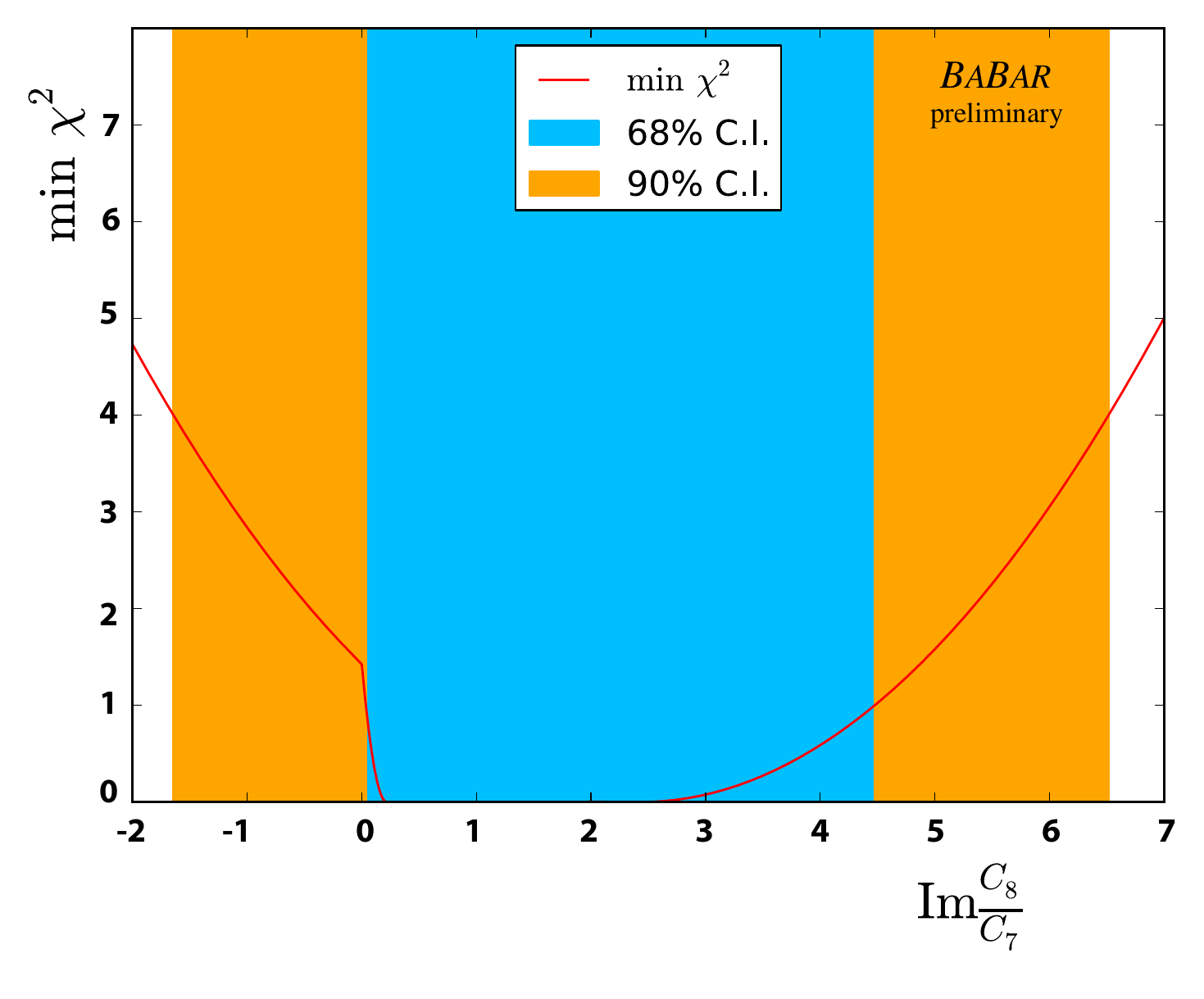}
\includegraphics[width=71mm]{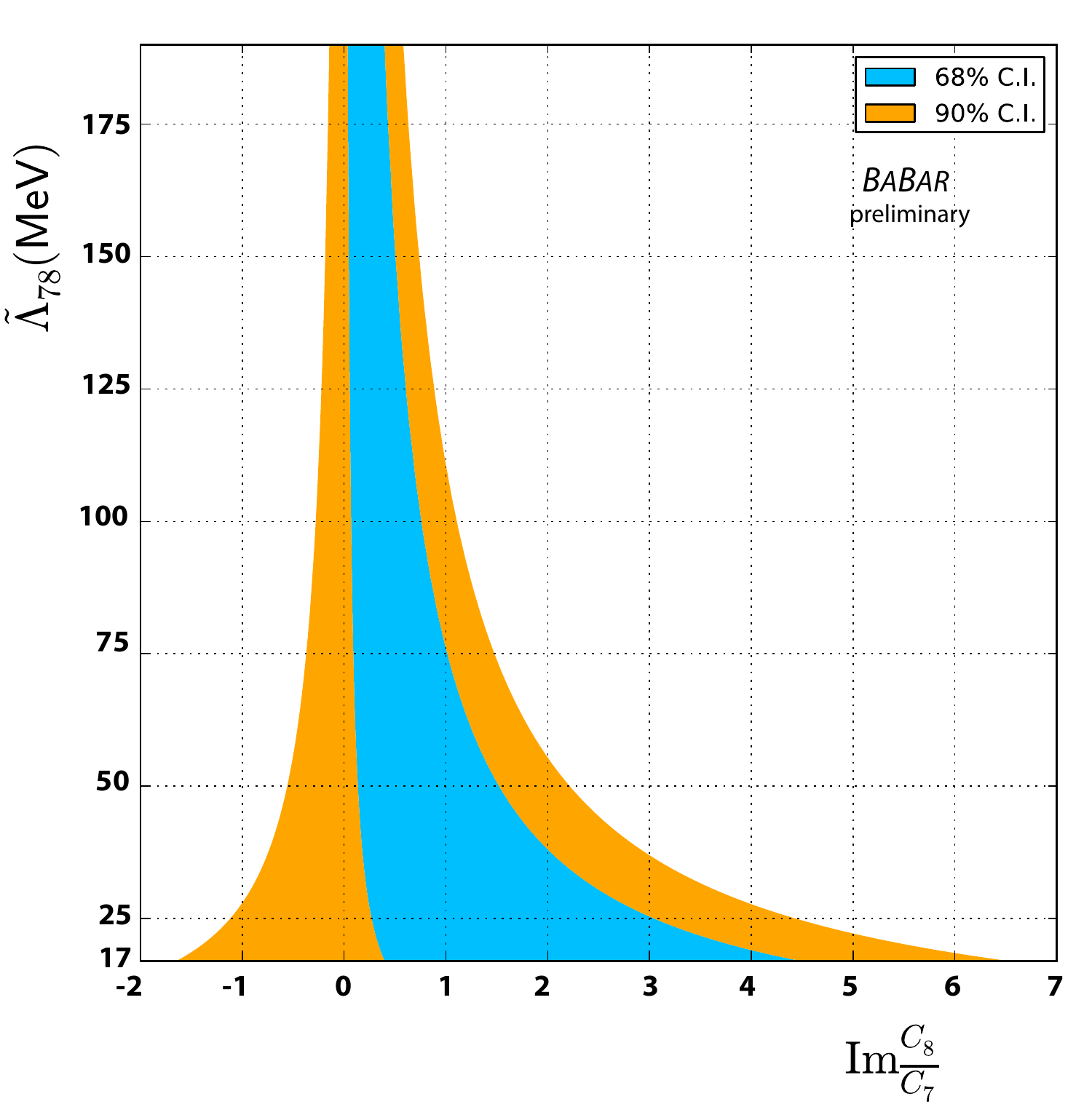}
\caption{The $\Delta \chi^2$ function versus ${\cal I}m(C^{eff}_8/C^{eff}_7)$ (left) and the dependence of $\bar \Lambda_{78}$ on ${\cal I}m(C^{eff}_8/C^{eff}_7)$ (right). The blue dark-shaded (orange light-shaded) regions show the $68\%~(90\%)$ CL intervals.}
 \label{fig:c78}
\end{figure}

Figure~\ref{fig:c78}
(left) show the $\Delta \chi^2$ of the simultaneous fit as a function of  ${\cal I}m(C^{eff}_8/C^{eff}_7)$. Figure~\ref{fig:c78} (right) shows the constraints of $\bar \Lambda_{78}$ as a function of ${\cal I}m(C^{eff}_8/C^{eff}_7)$. 
Tthe shape of $\Delta \chi^2$ as a function of  ${\cal I}m(C^{eff}_8/C^{eff}_7)$ is not parabolic indicating that the likelihood has a non-Gaussian shape. The reason is that $\Delta \chi^2$ is determined from all possible values of  $\bar \Lambda_{78}$. In the region $\sim 0.2 <  {\cal I}m(C^{eff}_8/C^{eff}_7) < \sim 2.6$ a change in 
${\cal I}m(C^{eff}_8/C^{eff}_7)$ $\Delta \chi^2$ can be compensated by a change in  $\bar \Lambda_{78}$ leaving 
 $\Delta \chi^2$ unchanged. For positive values larger (smaller) than 2.6 (0.2), $\Delta \chi^2$ increases slowly (rapidly), since  $\bar \Lambda_{78}$ remains nearly constant at the minimum value (increases rapidly). For negative ${\cal I}m(C^{eff}_8/C^{eff}_7)$ values, $\bar \Lambda_{78}$ starts to decrease again, which leads to a change in  $\Delta \chi^2$ shape.

\begin{figure}[h]
\centering
\includegraphics[width=120mm]{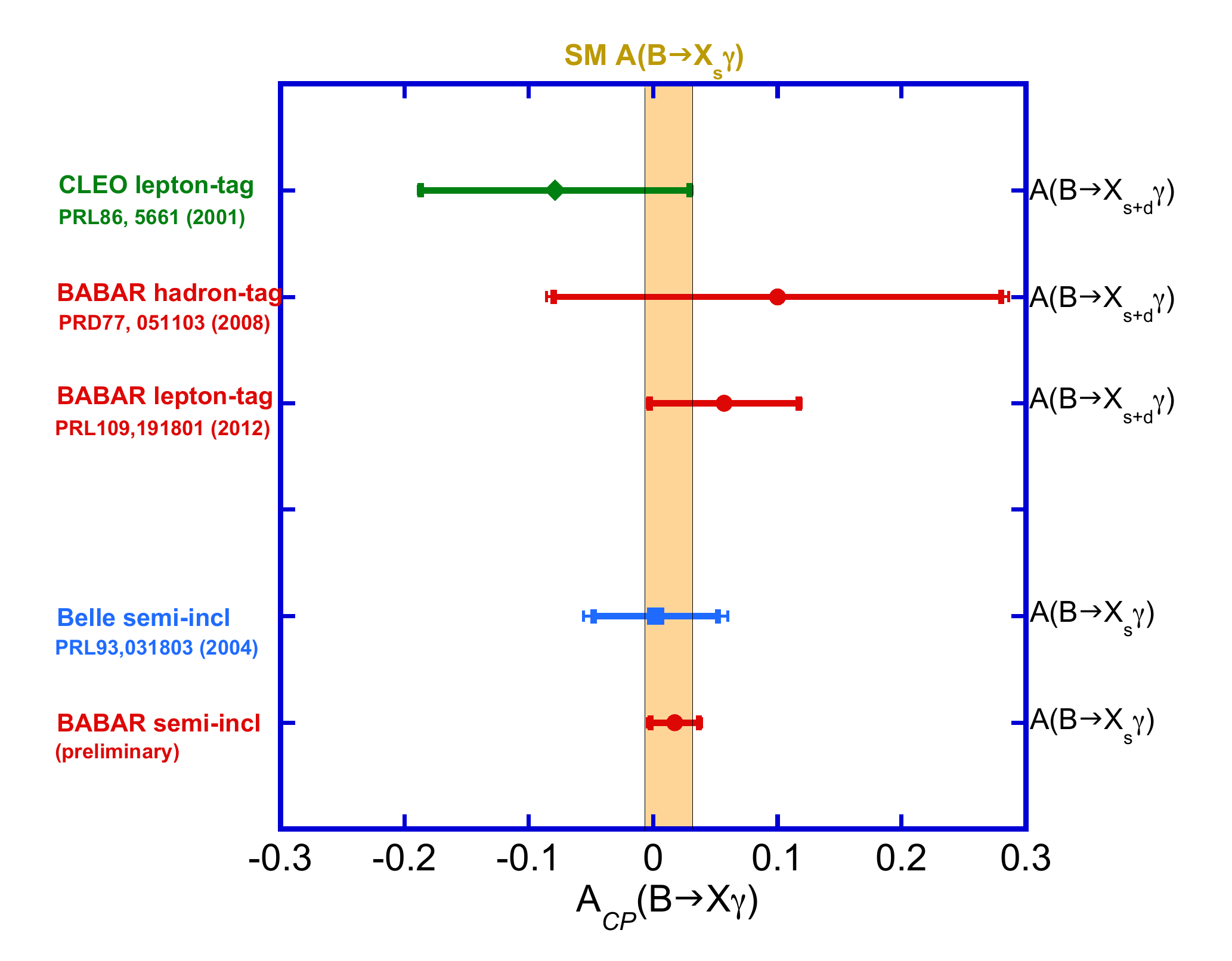}
\caption{Summary of  ${\cal A}_{C \! P}$ measurements for $B \ra X_s \gamma$ from semi-inclusive analyses~(\sbabar\ preliminary, Belle~\cite{belle4})  and for $B \ra X_{s+d} \gamma$ from fully inclusive analyses (\sbabar~\cite{babar12, babar8} and CLEO~\cite{cleo1a}) in comparison to the SM prediction for $B \ra X_s \gamma$~\cite{benzke}.}
 \label{fig:acpsg}
\end{figure}

In the fully inclusive analysis,  ${\cal A}_{C \! P}$ involves contributions from $B\ra X_s \gamma$ and $B\ra X_d \gamma$ that cannot be separated on an event-by-event basis. Therefore, we define  ${\cal A}_{C \! P}$ here as

\begin{equation}
{\cal A}_{CP} ( B \ra X_{s+d} \gamma) \equiv 
\frac{({\cal B}(\bar B  \ra X_{s+d} \gamma) - {\cal B} (B  \ra X_{s+d} \gamma) )}
        {({\cal B}(\bar B  \ra X_{s+d} \gamma) +{ \cal B}(B  \ra X_{s+d} \gamma) )}.
\end{equation}
\noindent 
We tag the $B$ flavor  by the lepton charge. Using a sample $384 \times 10^6~ B \bar B$ events, we measure
 ${\cal A}_{C \! P} (B \ra X_{s+d} \gamma) =0.057\pm 0.06_{stat} \pm 0.018_{sys}$ after correcting for charge bias and mistagging~\cite{babar12}. Figure~\ref{fig:acpsg} shows all  ${\cal A}_{C \! P}$ measurements from \sbabar~\cite{babar12, babar12b, babar8}, Belle~\cite{belle4} and CLEO~\cite{cleo1}. They all agree well with the SM prediction~\cite{kagan, hurth}. 
 
\section{$B \ra K^{(*)} \ell^+ \ell^-$ }

Using $471 ~(657) \times 10^6~ B\bar B$ events, \sbabar\  (Belle) reconstructs eight  (ten) $B \ra K^{(*)} \ell^+ \ell^-$ final states consisting of $K^+, K^0_S, K^+ \pi^-, K^0_S \pi^+, (K^+ \pi^0)$ recoiling against $e^+ e^-$ or $\mu^+ \mu^-$ \cite{babar12c, belle9a}. 
\sbabar\ (Belle) selects $e^\pm$ with momenta $p_e> 0.3~(0.4)~\rm GeV/c$. Both experiments select muons with
$p_\mu> 0.7~\rm GeV/c$, require good particle identification for $e^\pm, \mu^\pm, \pi^\pm $ and $K^\pm$ and  reconstruct $K^0_S$ in the $\pi^+ \pi^-$ final state.
To suppress combinatorial $q \bar q$ and $B \bar B$ backgrounds, \sbabar\ uses eight boosted decision trees (BDT),\footnote{two BDTs are used to separate signal from $B \bar B$ and $q \bar q$ backgrounds, separately for $e^+ e^-$ and $\mu^+ \mu^-$  modes and separately for $s$ below and above the $J/\psi$ mass.} while Belle uses likelihood ratios. Both experiments select signal with $m_{ES}$ and $\Delta E$ 
and veto the $J/\psi$ and $\psi(2S)$ mass regions. The vetoed  $J/\psi$ and $\psi(2S)$ samples and generated pseudo experiments are used to check the performance of the selection. To extract signal yields, both experiments perform one-dimensional fits of the $m_{ES}$ distributions for $B \ra K \ell^+ \ell^-$  modes and two-dimensional fits of the $m_{ES}$ and $m_{K\pi}$ mass distributions for $B \ra K^{*} \ell^+ \ell^-$ modes.

\begin{figure}[h]
\centering
\includegraphics[width=69mm]{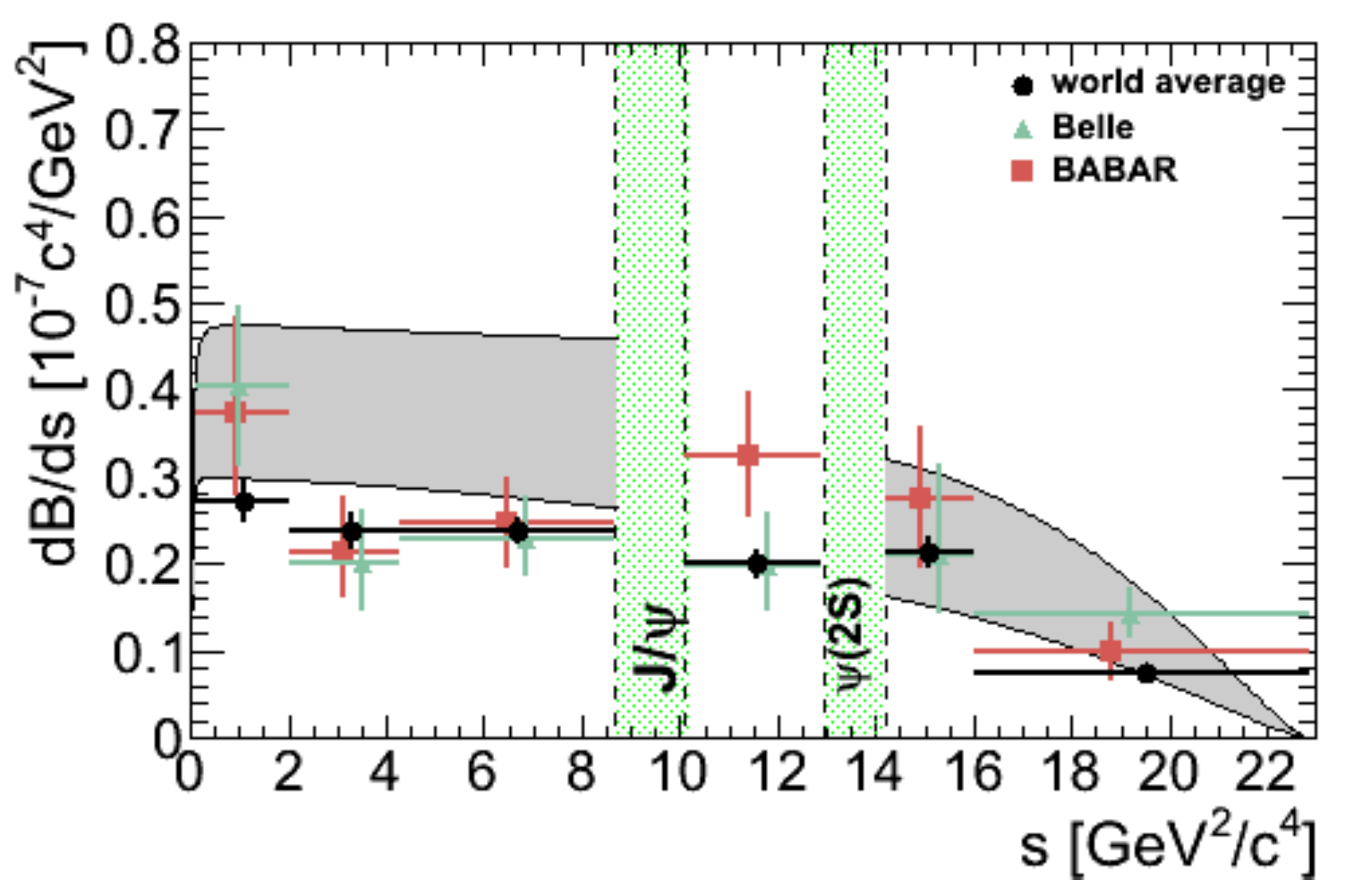}
\includegraphics[width=81mm]{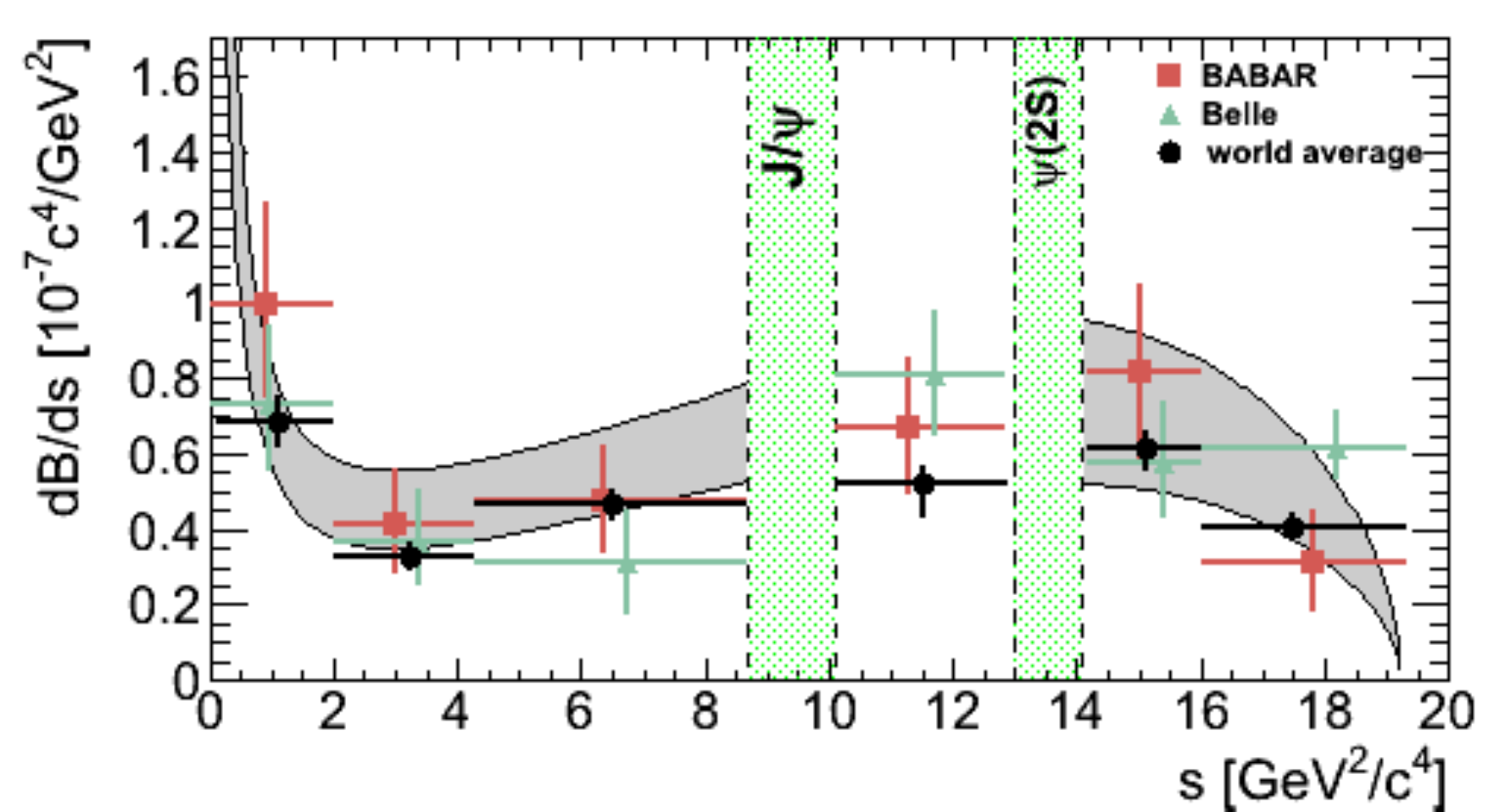}
\caption{$d {\cal B}/ds$ measurements for $B \ra K \ell^+ \ell^-$ (left) and $B\ra K^* \ell^+ \ell^-$ (right) from \sbabar~\cite{babar12c} (red squares), Belle~\cite{belle9a} (green triangles) and a naive world average (black points) that is dominated by LHCb in comparison to the SM predictions~\cite{bobeth12a} (grey curves). Vertical bands show the $J/\psi$ and $\psi(2S)$ vetoed regions.}
\label{fig:bsll}
\end{figure}

\subsection{$B \ra K^{(*)} \ell^+ \ell^-$ Rates and Rate Asymmetries}

\sbabar~\cite{babar12c} and Belle~\cite{belle9a} measured total and partial branching fractions $d {\cal B}(B \ra K^{(*)} \ell^+ \ell^-)/ds$ in six $s=q^2=m^2_{\ell^+ \ell^-}$ bins.\footnote{q is the momentum transfer and $m_{\ell^+ \ell^-}$ is the dilepton mass.} Figure~\ref{fig:bsll} (left) shows the \sbabar\ and Belle $d {\cal B}(B \ra K \ell^+ \ell^-)/ds$ measurements in comparison to a naive average that 
includes the $B \ra K \ell^+ \ell^-$ modes from \sbabar~\cite{babar12c} and Belle~\cite{belle9a} and $B \ra K \mu^+ \mu^-$ modes from CDF~\cite{cdf11} and LHCb~\cite{lhcb13}. Figure~\ref{fig:bsll} (right) shows the corresponding $d {\cal B}(B \ra K^* \ell^+ \ell^-)/ds$ results. The average is calculated using $B \ra K^* \ell^+ \ell^-$ modes from \sbabar~\cite{babar12c} and Belle~\cite{belle9a}, $B \ra K^* \mu^+ \mu^-$ modes from CDF~\cite{cdf11} and LHCb~\cite{lhcb12, lhcb13a} as well as the $B \ra K^{*0} \mu^+ \mu^-$ mode from CMS~\cite{cms12}. 
Note that the average values are dominated by the LHCb result. All measurements agree well with the SM predictions that are calculated for low and high values of $s$~\cite{bobeth10, bobeth12, bobeth12a}. For low $s$, the hadronic recoil is large and the $K^{(*)}$ energy is much larger than the QCD scale $\Lambda ~(E_{K^{(*)}} >> \Lambda)$. This region represents  the perturbative regime in which QCD factorization yields reliable results~\cite{beneke, beneke5}. For high $s\sim {\cal O}(m_b)$, the hadronic recoil becomes small and $E_{K^{(*)}} \sim \Lambda$. This is the non-perturbative regime in which an operator product expansion in powers of $1/m_b$ yields reliable results. The large uncertainties in the SM predictions result from the uncertainties in calculating the form factors of the hadronic matrix elements~\cite{ball}.

\begin{figure}[h]
\centering
\includegraphics[width=120mm]{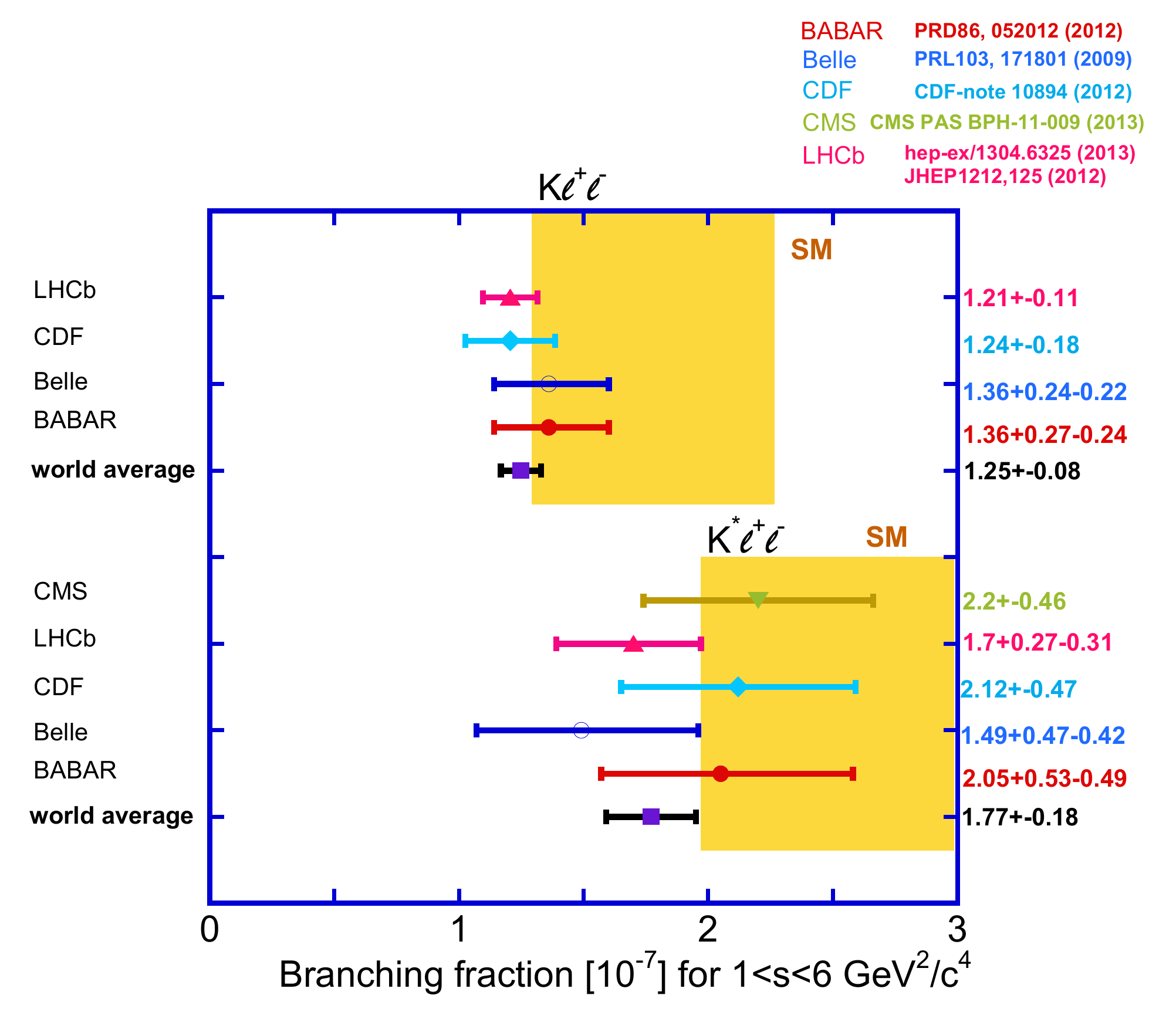}
\caption{Branching fraction measurements for $B \ra K \ell^+ \ell^-$  and $B\ra K^* \ell^+ \ell^-$ in the low $s$ region ($1 < s < 6~\rm GeV/c^2$) from \sbabar~\cite{babar12c}, Belle~\cite{belle9a}, CDF~\cite{cdf11}, LHCb~\cite{lhcb13, lhcb13a} and CMS~\cite{cms12} in comparison to the SM predictions~\cite{bobeth12a}.}
\label{fig:bsll-bflow}
\end{figure}

\begin{figure}[h]
\centering
\includegraphics[width=100mm]{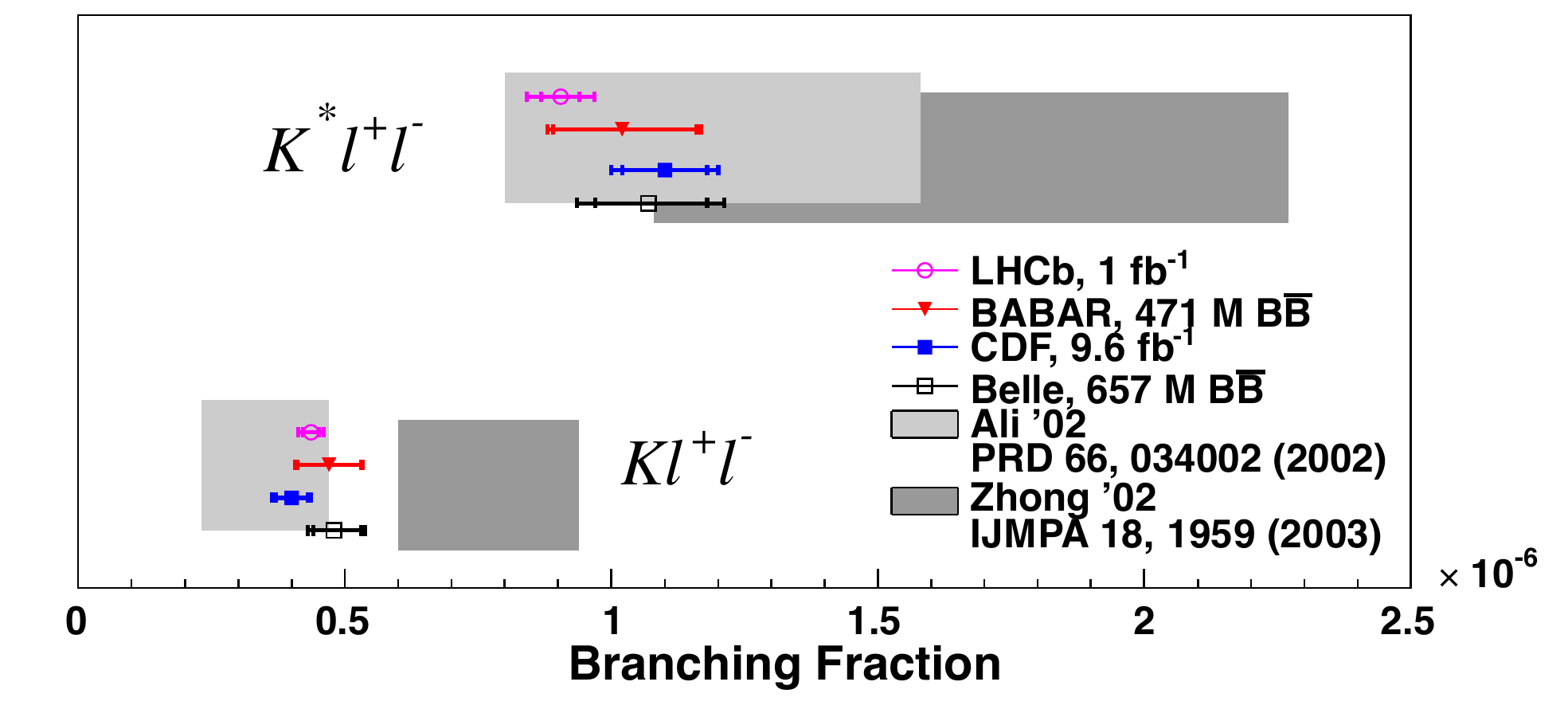}
\caption{Total branching fraction measurements for $B \ra K \ell^+ \ell^-$  and $B\ra K^* \ell^+ \ell^-$ from \sbabar~\cite{babar12b}, Belle~\cite{belle9a}, CDF~\cite{cdf11} and LHCb~\cite{lhcb13, lhcb13a} in comparison to the SM predictions~\cite{ali, zhong}.}
\label{fig:bsll-bftot}
\end{figure}

Figure~\ref{fig:bsll-bflow} shows all $B \ra K \ell^+ \ell^-$ and $B\ra K^* \ell^+ \ell^-$  branching fractions measured in the low $s$ region, $1 < s < 6 ~\rm GeV/c^2$, in comparison to SM predictions~\cite{bobeth12a}. Figure~\ref{fig:bsll-bftot} shows $B \ra K^{(*)} \ell^+ \ell^-$ total branching fraction measurements in comparison to SM predictions~\cite{ali, zhong}. All measurements show better agreement with the Ali model~\cite{ali}.
\sbabar\ measured total branching fractions of ${\cal B}(B \ra K \ell^+ \ell^-)=(4.7\pm0.6_{stat} \pm0.2_{sys}) \times 10^{-7}$ and ${\cal B}(B\ra K^* \ell^+ \ell^-)=(10.2^{+1.4}_{-1.3~stat}\pm 0.5_{sys})\times 10^{-7}$.
Table~\ref{tab:bsll} summarizes the \sbabar\  total branching fraction and rate asymmetry measurements.

The isospin asymmetry is defined by
\begin{equation}
d{\cal A}_I /ds \equiv \frac{d {\cal B} (B^0 \ra K^{(*)0} \ell^+ \ell^-)/ds - r_\tau d {\cal B} (B^+ \ra K^{(*)+} \ell^+ \ell^-)/ds}{d {\cal B} (B^0 \ra  K^{(*)0} \ell^+ \ell^-)/ds +r_\tau d {\cal B} (B^+ \ra K^{(*)+} \ell^+ \ell^-)/ds},
\end{equation}
\noindent
where  $r_\tau =\tau_{B^0}/ \tau_{B^+} $ accounts for the different $B^0$ and $B^+$ lifetimes. In the SM, ${\cal A}_I $ is expected to be at the order of ${\cal O}(1\%)$~\cite{matias03}. 

\begin{figure}[h]
\centering
\includegraphics[width=75mm]{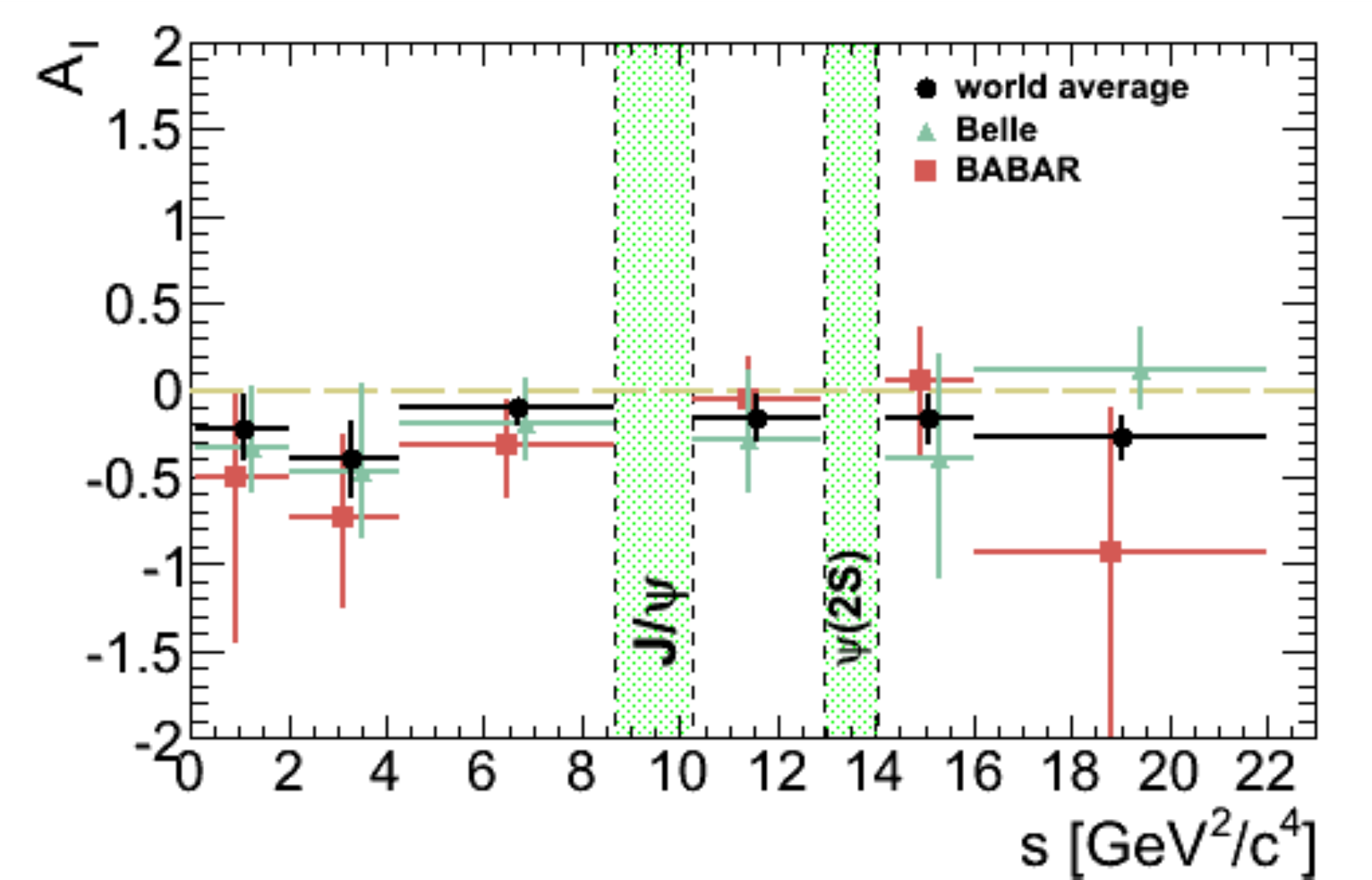}
\includegraphics[width=75mm]{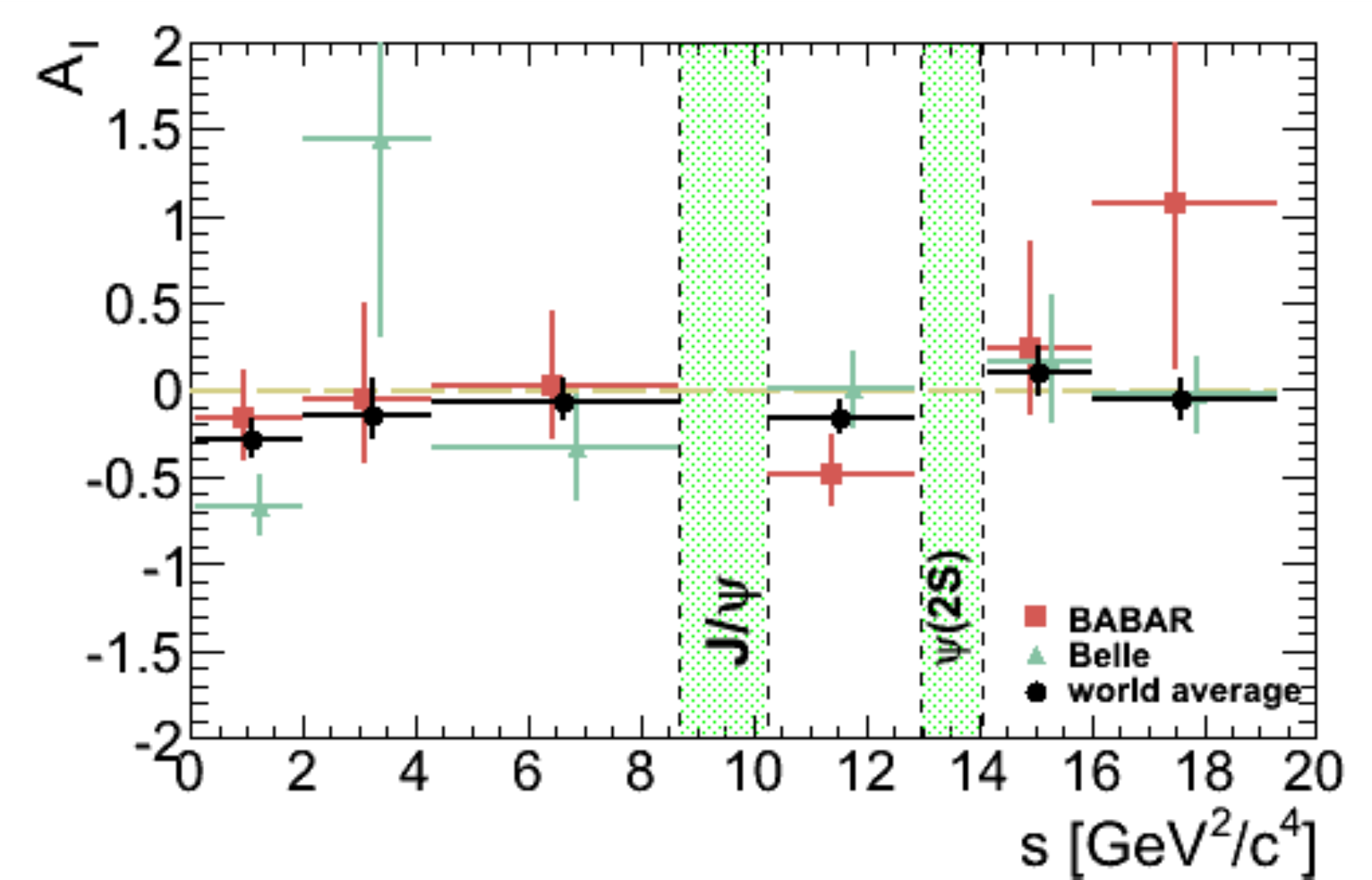}
\caption{Isospin asymmetry for $B \ra K \ell^+ \ell^-$ (left) and $B \ra K^* \ell^+ \ell^-$ (right) for \sbabar~\cite{babar12b} (red squares),
Belle~\cite{belle9a} (green triangles) and a naive world average over all experiments (black points) that is dominated by LHCb. Vertical bands show the  $J/\psi$ and $\psi(2S)$ vetoed regions. The dashed line indicates the SM prediction~\cite{matias03}.}
 \label{fig:ai}
\end{figure}

Figure~\ref{fig:ai} shows \sbabar\ and Belle isospin asymmetry measurements in six $s$ bins for $B \ra K \ell^+ \ell^-$ modes (left)  and $B \ra K^* \ell^+ \ell^-$ modes (right) in comparison to a naive average over all experiments~(\sbabar~\cite{babar12b}, Belle~\cite{belle9a}, CDF~\cite{cdf11}, and LHCb~\cite{lhcb12a}). The average points  are dominated again by LHCb. At low $s ~(1< s< 6~\rm GeV/c^2)$, the naive average yields ${\cal A}_I^{low~s} (B \ra K \ell^+ \ell^-)= -0.31\pm 0.12$ and  ${\cal A}_I^{low~s} (B \ra K^* \ell^+ \ell^-)= -0.15\pm 0.11$. 
For $B \ra K \ell^+ \ell^-$, consistency with the SM is at the $\sim 2.6 \sigma$ level. For other $s$ values and for $B \ra K^* \ell^+ \ell^-$, the averaged data agree well with the SM prediction~\cite{matias03}. The \sbabar\ measurements are listed in Table~\ref{tab:bsll}.

The  \CP asymmetry is defined by
\begin{equation}
\noindent
{\cal A}_{C \! P} = \frac{ {\cal B} (\bar B \ra \bar K^{(*)} \ell^+ \ell^-) -{\cal B} (B \ra K^{(*)} \ell^+ \ell^-)}{ {\cal B} (\bar B \ra \bar K^{(*)} \ell^+ \ell^-)+ {\cal B} (B \ra K^{(*)} \ell^+ \ell^-)}.
\end{equation}
\noindent
In the SM, the \CP asymmetry is expected to be small, ${\cal A}_{C \! P} =-0.01$~\cite{kruger00, bobeth08}. The measurements from \sbabar~\cite{babar12} (see Table~\ref{tab:bsll}), Belle~\cite{belle9} and LHCb~\cite{lhcb12b}  agree well with the SM prediction. 

The lepton flavor ratios are defined by

\begin{equation}
{\cal R}_{K^{(*)}} = {\cal B} (B \ra K^{(*)} \mu^+ \mu^-)/ {\cal B} (B \ra K^{(*)} e^+ e^-).
\end{equation}

\noindent
In the SM for  $s> 4 m_\mu^2$,\footnote{$m_\mu$ is the muon mass.} ${\cal R}_K^{(*)}\equiv 1$~\cite{ali00}. 
For  $s > 0.1~\rm GeV^2/c^4$, \sbabar~\cite{babar12} (see Table~\ref{tab:bsll}) and Belle~\cite{belle9} measure lepton flavor ratios that are consistent with unity and thus agree well with the SM prediction. 

Except for ${\cal A}_I(B \ra K \ell^+ \ell^-)$ at low $s$, all other measurements of branching  fractions and rate asymmetries are in good agreement with the SM predictions.

\begin{table}
\centering
\caption{\sbabar\ results for $B \ra K^{(*)} \ell^+ \ell^-$ modes on total branching fractions, \CP asymmetries, lepton flavor ratios and isospin asymmetries.
The first uncertainty is statistical, the second is systematic.}
\medskip
{\footnotesize
\begin{tabular}{|l|c|c|c|c|}
\hline \hline 
Mode & ${\cal B} [10^{-7}]$ & ${\cal A}_{CP}$ & ${\cal R}_{K^{(*)}}$ & ${\cal A}_I$  \T \B\\
 s $[\rm \frac{GeV^2}{c^4}]$ & all s & all s& $s > 0.1~\rm GeV^2/c^4$ & $0.1 \leq s \leq 8.12 \rm $ \T \B
\\ \hline
$K \ell^+ \ell^- $  &  $4.7\pm 0.6 \pm 0.2 $ 
& $-0.03\pm 0.14 \pm 0.01$ & $1.00 ^{+0.31}_{-0.25}\pm 0.07$  & 
$-0.58^{+0.29}_{-0.37}\pm 0.02$ \T \B \\
$K^* \ell^+ \ell^- $  &$10.2^{+ 1.4}_{-1.3}\pm 0.05$ & $\ 0.03 \pm 0.13 \pm 0.01$ & $ 1.13^{+0.34}_{-0.26}\pm 0.10$& $-0.25^{+0.17}_{-0.20}\pm 0.03$ \T \B  \\
\hline \hline                     
\end{tabular}
}
\label{tab:bsll}
\end{table}
\medskip

\subsection{$B \ra K^{(*)} \ell^+ \ell^-$ Angular Analyses}

The $B \ra K^{*} \ell^+ \ell^-$ decay is characterized by three angles: $\theta_K$ is the angle between the $K$ and $B$ in the $K^*$ rest frame, $\theta_\ell$ is the angle between the $\ell^+$ and the $B$ in the $\ell^+ \ell^-$ rest frame and $\phi$ is the angle between the $K^*$ and $\ell^+ \ell^-$ decay planes.  
The one-dimensional $\cos \theta_K$ and $\cos \theta_\ell$ projections depend on the  $K^*$ longitudinal polarization  ${\cal F}_L$ and the lepton forward-backward asymmetry ${\cal A}_{F \! B} $~\cite{ali00, bobeth07}

\begin{eqnarray}
W(\cos \theta_K) &=& \frac{3}{2} {\cal F}_L \cos^2 \theta_K + \frac{3}{4} ( 1- {\cal F}_L ) \sin^2 \theta_K,  \nonumber\\
W(\cos \theta_\ell) &=& \frac{3}{4} {\cal F}_L \sin^2 \theta_\ell +\frac{3}{8} (1- {\cal F}_L) (1+ \cos^2 \theta_\ell) +{\cal A}_{F \! B} \cos \theta_\ell.
\end{eqnarray}
\noindent
In the SM,  ${\cal A}_{F \! B} $ and ${\cal F}_L$ are again calculated separately for the low $s$ and high $s$ regions.

Since the number of signal events in each $s$ bin is small, \sbabar\ and Belle analyze one-dimensional angular distributions. Using $471~B \bar B$ events, \sbabar\ reconstructs six $B \ra K^{*} \ell^+ \ell^-$ final states with $K^* \ra K^+ \pi^-, K^0_S\pi^+, K^+ \pi^0$. The event selection is similar to that for rate asymmetries. \sbabar\ extracts ${\cal F}_L$ and ${\cal A}_{F \! B} $ by performing a profile likelihood scan. Using $657~B \bar B$ events, Belle performs a fit to the one-dimensional angular distributions.

 \begin{figure}[h]
\centering
\includegraphics[width=80mm]{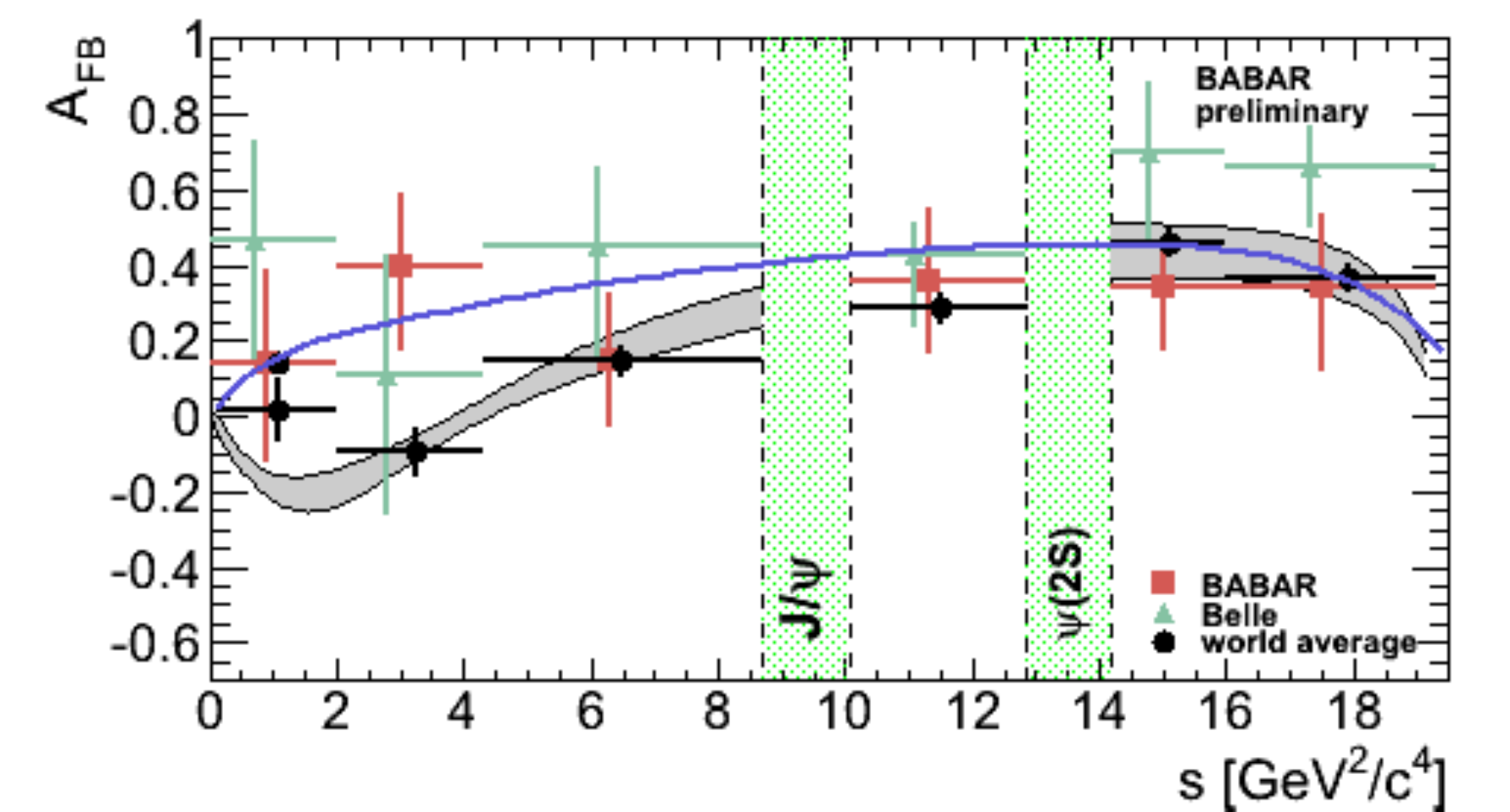}
\includegraphics[width=70mm]{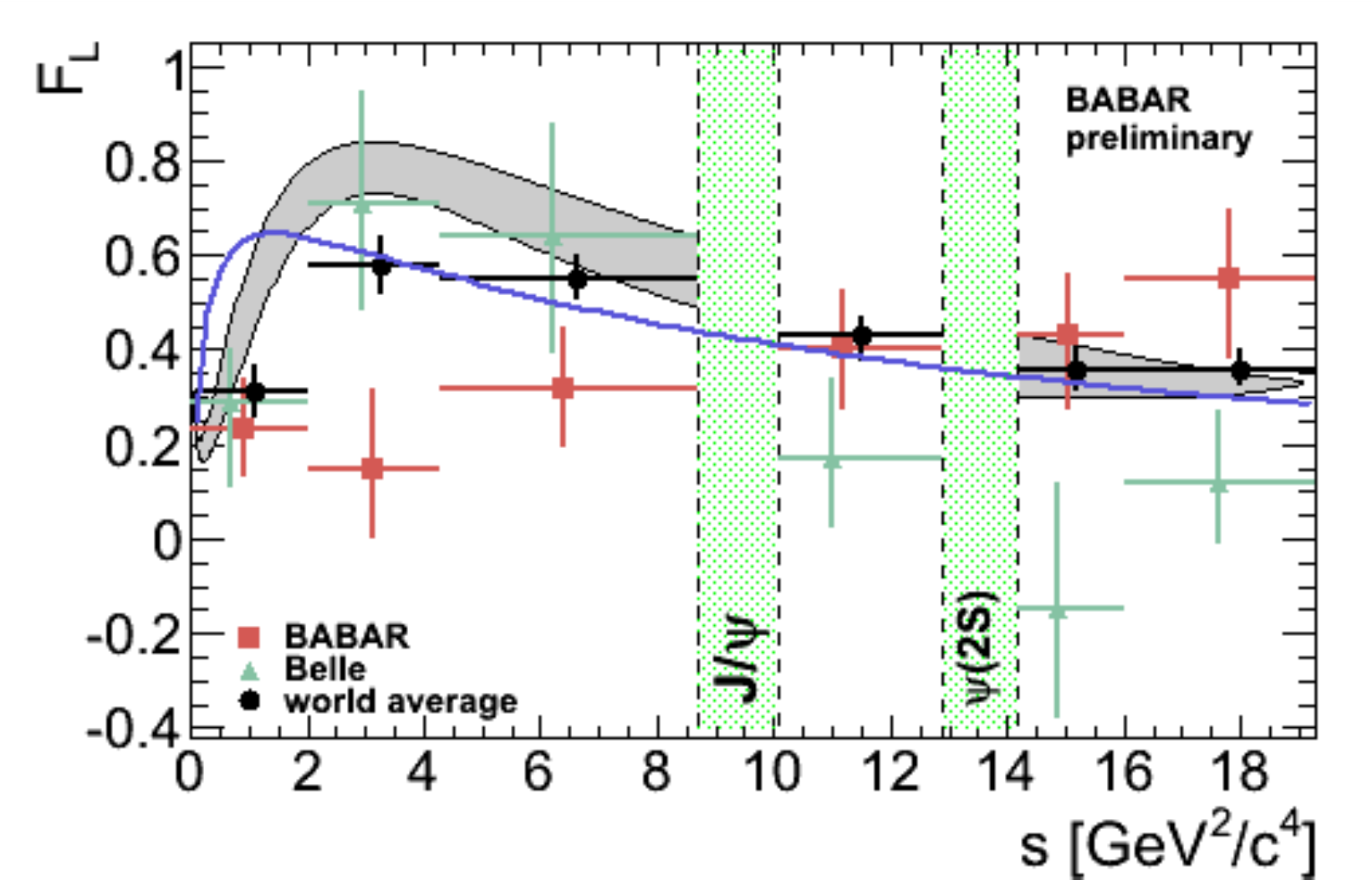}
\caption{\sbabar\ preliminary measurements (red squares) and Belle results~\cite{belle9a} (green triangles) for ${\cal A}_{F \! B}$ (left) and ${\cal F}_L$  (right) for $B\ra K^* \ell^+ \ell^-$modes in comparison to the naive world average over all experiments (black points) that is dominated by LHCb, the SM prediction (shaded curves) and a model in which the sign of \Cseven is flipped (blue solid curve). Vertical bands show the  $J/\psi$ and $\psi(2S)$ vetoed regions.}
 \label{fig:afb}
\end{figure}

Figure~\ref{fig:afb} (left) shows ${\cal A}_{F \! B} $ measurements in six $s$ bins from \sbabar\ (preliminary) and Belle~\cite{belle9a} in comparison to a naive average over the $B \ra K^* \ell^+ \ell^-$ results from \sbabar\ and Belle~\cite{belle9a}, $B \ra K^* \mu^+ \mu^-$ results from CDF~\cite{cdf11} and $B \ra K^{*0} \mu^+ \mu^-$ results from LHCb~\cite{lhcb13a}, CMS~\cite{cms12} and ATLAS~\cite{atlas}. The average values are dominated again by the LHCb measurements. In addition, predictions are shown for the SM and for a model in which the sign of Wilson coefficient \Cseven is flipped with respect to the expected value in the SM~\cite{ali, kruger00, ali00, buchalla}. The large uncertainties in the SM predictions result from uncertainties in the form factor calculations.
While the \sbabar\ measurements are consistent with both the SM and the flipped-sign \Cseven model, the world average values agree well with the SM prediction. In the low $s$ region, the world average yields ${\cal A}_{F \! B} (B \ra K^* \ell^+ \ell^-) =-0.074^{+0.047}_{-0.048}$, which agrees well with the SM prediction of ${\cal A}_{F \! B}^{SM} (B \ra K^* \ell^+ \ell^-) =-0.0494^{+0.0281}_{-0.0252}$ (for \sbabar\ results see Table~\ref{tab:afb}).

 Figure~\ref{fig:afb} (right) shows ${\cal F}_L$ measurements from \sbabar~(preliminary) and Belle~\cite{belle9a}
 in six $s$ bins in comparison to a naive average using $B \ra K^* \ell^+ \ell^-$ results from \sbabar\ and Belle~\cite{belle9a}, $B \ra K^* \mu^+ \mu^-$ results from CDF~\cite{cdf11} and $B \ra K^{*0} \mu^+ \mu^-$ results from LHCb~\cite{lhcb13a}, CMS~\cite{cms12} and ATLAS~\cite{atlas}. The naive average values are again dominated by the LHCb results. The figure also shows the SM prediction~\cite{bobeth12a} and the prediction of the flipped-sign \Cseven model~\cite{ali, kruger00, ali00}.  All results are consistent with the SM prediction, though the \sbabar\ results fit better to the flipped-sign \Cseven model.  In the low $s$ region ($1 < s < 6~\rm GeV^2/c^4$), the world average yields ${\cal F}_L (B \ra K^* \ell^+ \ell^-) =0.523^{+0.047}_{-0.044}$, which  is consistent with the SM prediction of ${\cal F}_L^{SM} (B \ra K^* \ell^+ \ell^-) =0.735^{+0.06}_{-0.07}$~\cite{bobeth12a, kruger00, ali00,  kruger05, buchalla}  (for \sbabar\ results see Table~\ref{tab:afb}).

\begin{table}
\centering
\caption{\sbabar\ measurements of the lepton forward-backward asymmetry and $K^*$ longitudinal polarization for $B \ra K^{*} \ell^+ \ell^-$ modes in the low $s$ region.
The first uncertainty is statistical, the second is systematic.}
\medskip
{\footnotesize
\begin{tabular}{|l|c|c|}
\hline \hline 
Mode &  ${\cal A}_{F \! B}$ & ${\cal F}_L$ \T \B \\
 s $[\rm \frac{GeV^2}{c^4}]$ &  $1.0 \leq s \leq 6.0\rm $ &  $1.0 \leq s \leq 6.0 $ \T \B  \\ \hline
$K^* \ell^+ \ell^- $  & $0.26^{+0.27}_{-0.30}\pm 0.07 $& $0.25^{+0.09}_{-0.08} \pm 0.03$ \T \B
  \\
\hline \hline                     
\end{tabular}
}
\label{tab:afb}
\end{table}

\section{Search for $B \ra \pi  \ell^+ \ell^-$ and $B \ra \eta  \ell^+ \ell^-$ Decays}

In the SM in lowest order, $ B \ra X_d \ell^+ \ell^-$ modes are also mediated by the electromagnetic penguin, $Z$ penguin and $WW$ box diagrams. However, they are suppressed by $|V_{td}/V_{ts}|^2 \sim 0.04$ with respect to the corresponding $B \ra X_s \ell^+\ell^-$ modes. In extensions of the SM, rates may increase significantly~\cite{aliev}. Using $471 \times 10^6 ~B \bar B$ events, \sbabar\  recently updated the search for $B \ra \pi \ell^+ \ell^-$ modes and performed the first search for $B \ra \eta \ell^+ \ell^-$ modes. The SM predictions lie in the range ${\cal B}(B \ra \pi \ell^+ \ell^-)= (1.96-3.30) \times 10^{-8}$ and ${\cal B}(B \ra \eta \ell^+ \ell^-)= (2.5-3.7) \times 10^{-8}$  where the large uncertainties result from uncertainties in the $B \ra \pi$ form factor calculations~\cite{aliev, wang, song} and from a lack of knowledge of $B \ra \eta$ form factors~\cite{erkol}.

\sbabar\ fully reconstructs four  $B \ra \pi\ell^+ \ell^-$ and four $B \ra \eta \ell^+ \ell^-$  final states by selecting $\pi^\pm, \pi^0 , \eta \ra \gamma \gamma$ and $\eta \ra \pi^+ \pi^- \pi^0$ recoiling against $e^+ e^-$ or $\mu^+ \mu^-$~\cite{babar13b}. We select  leptons with $p_\ell > 0.3~\rm GeV/c$, recover losses due to bremsstrahlung for $e^\pm$, remove $\gamma \ra e^+ e^-$ decays and require good particle identification for $e^\pm, \mu^\pm$ and $\pi^\pm$. We select photons with $E_\gamma > 50 ~\rm MeV$ and impose a $\pi^0$ mass constraint of $115 < m_{\gamma \gamma} < 150~\rm MeV/c^2$  and an $\eta$ mass constraint of $ 500 ~(535) < m_{\gamma \gamma}~ (m_{3\pi}) < 575 ~(565) ~\rm MeV/c^2$. In addition, we require $(E_{1,\gamma} - E_{2,\gamma} )/(E_{1,\gamma} + E_{2,\gamma} ) <0.8$ for the $\eta \ra \gamma \gamma$ final states to remove asymmetric $q \bar q$ background that peaks near one. We veto $J/\psi$ and $\psi(2S)$ mass regions and use four NNs to suppress combinatorial $B \bar B$ and $q \bar q$ continuum backgrounds, separately for $e^+ e^-$ modes and for  $\mu^+ \mu^-$ modes.  The NNs for suppressing $B \bar B $ background uses 15 (14) input distributions for $e^+ e^-~ (\mu^+ \mu^-)$ modes, while those for suppressing $q \bar q$ continuum use 16  input distributions for both modes.  For validations, we use pseudo-experiments and the vetoed $J/\psi$ and $\psi(2S)$ samples. 
The selection criteria used by Belle are given in ~\cite{belle8}.

For $B \ra \pi^+  \ell^+ \ell^-$ and $B \ra \pi^0  \ell^+ \ell^-$, \sbabar\ performs simultaneous unbinned maximum likelihood fits to the $m_{ES}$ and $\Delta E$ distributions for $e^+ e^-$ and $\mu^+ \mu^-$ modes separately.
We include the $B \ra K^+  \ell^+ \ell^-$ mode in the fit  to extract the peaking background contribution in the $B \ra \pi^+ \ell^+ \ell^-$ modes by reconstructing the $K^+$ as a $\pi^+$. We use the vetoed $J/\psi$ and $\psi(2S)$ samples to validate the fit and check the peaking  $B \ra K^+  \ell^+ \ell^-$ contribution.

For the $B \ra \eta  \ell^+ \ell^-$, we perform simultaneous unbinned maximum likelihood fits to the $m_{ES}$ and $\Delta E$ distributions, again  for $e^+ e^-$ and $\mu^+ \mu^-$ modes separately. We use the vetoed $J/\psi$ and $\psi(2S)$ samples to validate the fits. In addition, we perform fits for the isospin-averaged modes $B\ra \pi e^+e^-$ and $B\ra \pi \mu^+\mu^-$, lepton-flavor averaged modes $B^+ \ra \pi^+ \ell^+\ell^-, B^0 \ra \pi^0 \ell^+\ell^-$ and $B^0 \ra \eta \ell^+\ell^-$ and both isospin and lepton-flavor averaged modes $B\ra \pi \ell^+\ell^-$.

\begin{figure}[h]
\centering
\includegraphics[width=110mm]{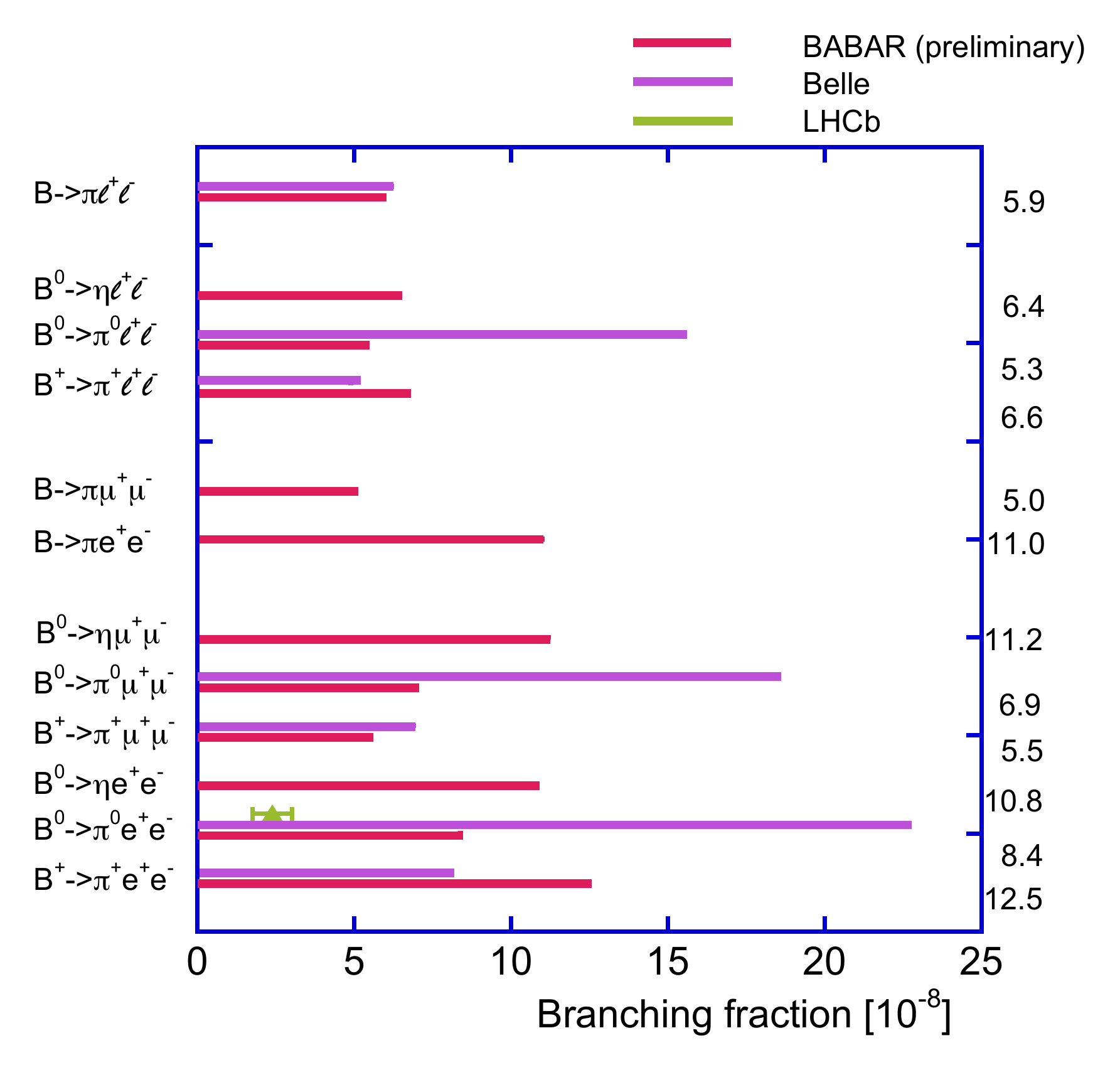}
\caption{Branching fraction upper limits at $90\%$ CL for $B \ra \pi \ell^+ \ell^-$ and $B \ra \eta \ell^+ \ell^-$ modes from \sbabar\ (preliminary) and Belle~\cite{belle8} and the measurement of  $B \ra \pi^+ \ell^+ \ell^-$ by LHCb~\cite{lhcb12c}.}
 \label{fig:dll}
\end{figure}

Similar to Belle, we see no signals in any of these modes and set branching fraction upper limits at $90\%$ CL. Recently, LHCb observed the $B \ra \pi^+ \ell^+ \ell^-$  decay and measured a branching fraction of ${\cal B} (B \ra \pi^+ \ell^+ \ell^-)=(2.4\pm 0.6 \pm 0.1) \times 10^{-8}$~\cite{lhcb12c}.  Figure~\ref{fig:dll} shows the preliminary \sbabar\ branching fraction upper limits in comparison to those from Belle~\cite{belle8} and the $B \ra \pi^+ \ell^+ \ell^-$ measurement from LHCb~\cite{lhcb12c}. \sbabar\ sets the best branching fraction upper limit for $B^0\ra \pi^0 \ell^+\ell^-$ and presents the first results for $B \ra \eta \ell^+\ell^-$ modes. Note that the present branching fraction upper limits lie within a factor of two to three of the SM predictions.

\section{Conclusion}

The \sbabar\   $B \ra X_s \gamma$ measurements of branching fractions, photon energy moments, $m_b$, $\mu_\pi^2$  are in good agreement with the SM predictions. The ${\cal B}(B  \ra X_s \gamma)$ measurement  provides a constraint on the charged Higgs mass of  $M_{H^\pm} > 327 ~\rm GeV/c^2$ at $95\% ~CL$ independent of $\tan \beta$. The new ${\cal A}_{C \! P} (B \ra X_s \gamma)$ measurement is the most precise result and agrees well with the SM prediction. The \sbabar\ $\Delta {\cal A}_{C \! P} (B \ra X_s \gamma)$ measurement is the first one and provides the first constraint on ${\cal I}m(C^{eff}_8/C^{eff}_7)$.  \sbabar\ updated branching fraction upper limits for $B \ra \pi \ell^+ \ell^-$ and presented the first branching fraction upper limits for $B \ra \eta \ell^+ \ell^-$.  

For $B \ra K^{(*)} \ell^+ \ell^-$, the measurements of branching fractions, isospin asymmetries, lepton flavor ratios, \CP asymmetries, $K^*$ longitudinal polarization and lepton forward-backward asymmetry averaged over all experiments agree with the SM predictions. The largest deviation from the SM prediction is less than $3\sigma$ and results from the isospin asymmetry of $B \ra K \ell^+ \ell^-$ in the low $s$ region. To look for more deviations from the SM, the precision of all measurements needs to be improved significantly.  Such improvements are expected to come from LHCb and Belle II. Furthermore, large data samples in these experiments will permit studies of the full angular distribution in $B \ra K^* \ell^+ \ell^-$, which is described by 12 observables~\cite{altmannshofer, hurth10}. 
Some the new observables have a higher discrimination power between the SM and new physics effects 
${\cal A}_{F \! B}$ and ${\cal F}_L$.



\bigskip
\bigskip

\Acknowledgements

I would like to thank members of the \sbabar\ collaboration for giving me the opportunity to present these results. In particular, I would like to thank Piti Ongmongkolkul, Bill Dunwoodie, David Hitlin and Jack Ritchie for their fruitful suggestions. Furthermore, I would like to thank Gudrun Hiller and Aoife Bharucha for supplying Mathematica files with the SM predictions for $B \ra K^{(*)} \ell^+ \ell^-$ modes.


\end{document}